\documentclass[prd,superscriptaddress,a4paper,showpacs,showkeys,10pt,nofootinbib]{revtex4}
\usepackage{graphicx}
\usepackage{graphics}
\usepackage{epsfig}
\usepackage{dcolumn}
\usepackage{bm}
\usepackage{multirow}
\usepackage{tabularx}
\usepackage{hyperref}
\usepackage{commath}
\usepackage{epstopdf}
\usepackage[T1]{fontenc}
\usepackage{geometry}
\usepackage{color}

\def\lsim{\raise0.3ex\hbox{$<$\kern-0.75em\raise-1.1ex\hbox{$\sim$}}}

\def\gsim{\raise0.3ex\hbox{$>$\kern-0.75em\raise-1.1ex\hbox{$\sim$}}}

\def\pom{{I\!\!P}}

\newcommand{\be}{\begin{equation}}

\newcommand{\ee}{\end{equation}}

\def\beq{\begin{equation}}

\def\eeq{\end{equation}}

\def\beqa{\begin{eqnarray}}

\def\eeqa{\end{eqnarray}}

\newcommand{\ba}{\begin{eqnarray}}

\newcommand{\ea}{\end{eqnarray}}

\newcommand{\rr}{\mbox{\boldmath $r$}}

\newcommand{\rb}{\mbox{\boldmath $b$}}

\def\gappeq{\mathrel{\rlap {\raise.5ex\hbox{$>$}}

{\lower.5ex\hbox{$\sim$}}}}

\def\lappeq{\mathrel{\rlap{\raise.5ex\hbox{$<$}}

{\lower.5ex\hbox{$\sim$}}}}

\def\Toprel#1\over#2{\mathrel{\mathop{#2}\limits^{#1}}}

\def\pom{{I\!\!P}}

\def\bi{\begin{itemize}}
\def\ei{\end{itemize}}

\def\be{\begin{eqnarray}}
\def\ee{\end{eqnarray}}
\def\bc{\begin{center}}
\def\ec{\end{center}}
\def\beq{\begin{equation}}
\def\eeq{\end{equation}}

\begin{document}

\title{Diffractive deeply inelastic scattering in future electron-ion colliders}

\author{D. Bendova}
\email[]{Dagmar.Bendova@fjfi.cvut.cz}
\affiliation{
Faculty of Nuclear Sciences and Physical Engineering, Czech Technical University in Prague,\\ B\v rehov\'a 7, 11519
Prague, Czech Republic \\}

\author{J. Cepila}
\email[]{jan.cepila@fjfi.cvut.cz}
\affiliation{
Faculty of Nuclear Sciences and Physical Engineering, Czech Technical University in Prague,\\ B\v rehov\'a 7, 11519
Prague, Czech Republic \\}

\author{J. G. Contreras}
\email[]{jesus.guillermo.contreras.nuno@cern.ch }
\affiliation{
Faculty of Nuclear Sciences and Physical Engineering, Czech Technical University in Prague,\\ B\v rehov\'a 7, 11519
Prague, Czech Republic \\}

\author{V. P. Gon\c{c}alves}
\email[]{barros@ufpel.edu.br}
\affiliation{High and Medium Energy Group, Instituto de F\'{\i}sica e Matem\'atica,  Universidade Federal de Pelotas (UFPel)\\
Caixa Postal 354,  96010-900, Pelotas, RS, Brazil. \\
}

\author{M. Matas}
\email[]{matas.marek1@gmail.com }
\affiliation{
Faculty of Nuclear Sciences and Physical Engineering, Czech Technical University in Prague,\\ B\v rehov\'a 7, 11519
Prague, Czech Republic \\}

\begin{abstract}
The impact of  nonlinear effects in the diffractive observables that will be measured in future electron-ion collisions is investigated. We present, for the first time, the  predictions for the diffractive structure function and reduced cross sections derived using the solution to the Balitsky--Kovchegov equation with the collinearly-improved kernel and including the impact-parameter dependence. We demonstrate that the contribution of the diffractive events is enhanced in nuclear collisions and that the study of the ratio between the nuclear and proton predictions will be useful to discriminate among different models of the dipole-target scattering amplitude and, consequently, will allow us to constrain the description of QCD dynamics in parton densities.
\end{abstract}



\maketitle

\vspace{1cm}

\section{Introduction}

The understanding of the high-energy (small-$x$) regime of quantum chromodynamics (QCD) is one of the main challenges of this theory~\cite{hdqcd}. Experimentally, this regime was intensely investigated in $ep$ collisions at HERA (DESY) and has been studied in $pp$, $pA$, and $AA$ collisions at RHIC (BNL) and at the LHC (CERN). These experiments indicate that gluons play a dominant role in the structure of hadrons, with the gluon density rapidly increasing at smaller values of $x$. Theoretically, the growth of the gluon distribution is expected to saturate, with the system forming a Color Glass Condensate (CGC), whose evolution with energy is described by an infinite hierarchy of coupled equations for the correlators of Wilson lines~\cite{CGC,BAL}. In the mean-field approximation, the first equation of this hierarchy decouples and boils down to a single non-linear integro-differential equation: the Balitsky-Kovchegov (BK) equation~\cite{BAL,kov}. Such equation determines, in the large-$N_c$ limit, where $N_c$ is the number of colors, the evolution of the two-point correlation function, which corresponds to the scattering amplitude ${\cal{N}}(x,\rr,\rb_t)$ of a dipole off the CGC, where $r = |\rr|$ is the transverse dipole size and $\rb_t$ the impact parameter. This quantity encodes the information about the hadronic scattering as well as the nonlinear and quantum effects in the hadron wave function. For recent reviews, see e.g.~\cite{hdqcd}. 

During recent years, the CGC formalism has been developed at higher accuracy and successfully applied to describe a large set of observables in $ep$, $pp$, $pA$, and $AA$ collisions. Although these results are very promising,  there is no clear consensus on whether the onset of the nonlinear regime has been reached. The search for these nonlinear effects is one of the major motivations for the construction of the Electron-Ion Collider (EIC) in the US~\cite{eic}, recently approved, as well as for the proposal of future electron-hadron colliders at CERN~\cite{lhec}. These colliders are expected to allow for the investigation of the hadronic structure with an unprecedented precision in inclusive and diffractive observables. In particular, electron-nucleus collisions are considered ideal to probe the nonlinear regime~\cite{eic_general}. The larger parton densities in the nuclear case, with respect to the proton case, enhance by a factor $\propto A^{1/3}$ the nuclear saturation scale, $Q^2_{s,A}$, which determines the onset of nonlinear effects in QCD dynamics. Such expectations have motivated an intense interest in phenomenology regarding the implications of gluon saturation effects in QCD observables~\cite{eic,lhec}. These studies demonstrated that the analysis of diffractive events can be considered a smoking gun of gluon saturation effects in $eA$ collisions \cite{Kowalski_prl,erike_ea2,vmprc,Caldwell,Lappi_inc,Toll,armestoamir,diego,Lappi:2014foa,Mantysaari:2016ykx,Mantysaari:2016jaz,Diego1,contreras,Luszczak:2017dwf,Mantysaari:2017slo,Diego2,Bendova:2018bbb,cepila,Lomnitz:2018juf,Mantysaari:2019jhh}. In particular, diffractive events are predicted to contribute with half of the total cross section in the asymptotic limit of very high energies, with the other half being formed by all inelastic processes~\cite{Nikolaev,simone2,Nik_schafer,Kowalski_prc}. In addition, the associated observables depend on the square of the scattering amplitude, which makes them  strongly sensitive to the underlying QCD dynamics. These results strongly motivate the study of diffraction in $eA$ collisions using as input in the calculations a realistic model for the scattering amplitude.

In this paper, we investigate the impact of nonlinear effects on diffractive observables that can be measured in future electron-hadron colliders. In particular, we  predict the diffractive cross section and diffractive structure functions using the color-dipole formalism and different models for the dipole-hadron scattering amplitude considering $ep$ and $eA$ collisions. Our goal is to improve the studies performed in Refs.~\cite{simone2,Kowalski_prl,erike_ea2,Kowalski_prc,GBW,vicmarcos,Marquet} in the following aspects: ($i$) unlike Refs.~\cite{simone2,GBW,vicmarcos}, we do not assume that the impact-parameter dependence of $\cal{N}$ can be factorized as ${\cal{N}}(x,\rr,\rb_t) = {{N}}(x,\rr) S(\rb_t)$, where $S(\rb_t)$ is the target profile function; ($ii$) the impact-parameter dependence of ${\cal{N}}(x,\rr,\rb_t)$ is derived using the BK equation and taking into account the collinear corrections to the kernel of the evolution equation, following the approach proposed in Refs.~\cite{cepila_nucleon,cepila_nucleon1} (in  Refs.~\cite{Kowalski_prc,Marquet} the $\rb$ dependence of ${\cal{N}}$ is an assumption of the phenomenological models considered); ($iii$) the predictions for $eA$ collisions are derived using the solution of the BK equation for nuclear targets obtained in Ref.~\cite{cepila_nuclear}, which takes into account the collinear corrections and the impact-parameter dependence, instead of the Glauber-Gribov (GG) approach~\cite{glauber,gribov,mueller} used in Refs.~\cite{Kowalski_prl, Kowalski_prc,erike_ea2}. For completeness of our study, we compare our predictions  with those derived using the IP-Sat and b-CGC models \cite{kmw,Kowalski:2003hm} for the dipole-proton scattering amplitude, generalized for the nuclear case using the GG approach.

The paper is organized as follows. In the next section, we present a brief overview of the color dipole formalism for the description of diffractive deeply inelastic scattering in $ep$ and $eA$ collisions. The impact-parameter dependent BK equation is presented and its solutions for the cases of a proton and a nucleus are discussed. In Sec.~\ref{sec:res}, we present our predictions for the ratio between the diffractive and the total cross sections as well as for the diffractive structure functions. A comparison to HERA data is presented and the nuclear dependence of the individual components of the diffractive structure functions is discussed in detail. Finally, in Sec.~\ref{sec:conc} we summarize our main conclusions.

 \begin{figure}[t]
 {\includegraphics[width=0.55\textwidth]{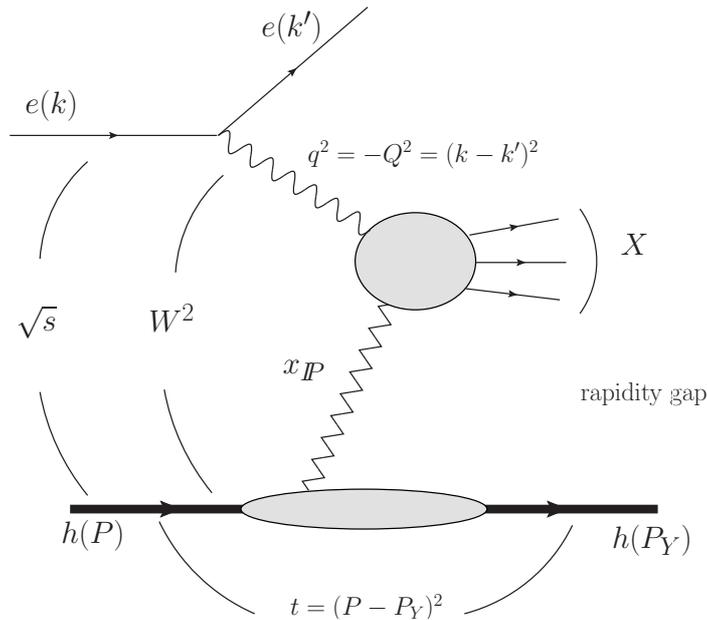}}  
\caption{Diffractive deeply inelastic scattering in electron-hadron collisions, where the hadron can be in particular a proton or a nucleus.}
\label{fig:ddis}
\end{figure}

\section{Formalism}

Diffractive electron-hadron scattering, $e h \rightarrow e X h$, is represented in Fig.~\ref{fig:ddis}, where the hadron in the final state carries most of the beam momentum and $X$ represents all the other final state particles. The basic idea is that in the $\gamma^* h$  interaction, the hadron remains intact and a  hadronic system $X$ with mass $M_X$ is produced with a rapidity gap between them. The fractional longitudinal momentum loss of the hadron is denoted as $x_{\pom}$ and is related to the photon virtuality $Q^2$ and $M_X^2$ as
\beq x_{\pom} = \frac{Q^2+M_{X}^2}{Q^2+W^2},\eeq
where $W$ is the photon-hadron center-of-mass energy. In addition, we can define the variable $\beta$ given by
\beq \beta = \frac{Q^2}{Q^2+M_X^2}, \eeq
which is related to the Bjorken variable $x = Q^2 / (Q^2+W^2)$ and $x_{\pom}$  by $x = \beta x_{\pom}$. The measured diffractive cross section can be expressed as
\begin{eqnarray}
\frac{\mathrm{d} \sigma^{eh \rightarrow eXh}}{\mathrm{d}\beta \mathrm{d}Q^2 \mathrm{d}x_{\pom}} = \frac{4\pi\alpha_{em}^2}{\beta Q^4}\left[1 - y + \frac{y^2}{2}\right]\,\sigma^{D(3)}_{r}(x_{\pom},\beta,Q^{2}) \, , 
\end{eqnarray}
where $y$ is the fractional energy loss of the electron in the hadron rest frame and the reduced cross section is related to the diffractive structure functions as
\begin{equation}
\sigma^{D(3)}_{r}(x_{\pom},\beta,Q^{2})=F^{D(3)}_{2}(x_{\pom},\beta,Q^{2})-\frac{y^{2}}{1+(1-y)^{2}}F^{D(3)}_{L}(x_{\pom},\beta,Q^{2}) \,.
\end{equation}
Experimentally, diffractive $eh$ scattering is characterized by the presence of a leading hadron at beam rapidities in the final state, and by a rapidity gap in between this hadron and the produced system $X$. The size of the rapidity gap is $\ln(1/x_{\pom})$.

 \begin{figure}[t]
\begin{tabular}{cc}
 {\includegraphics[width=0.5\textwidth]{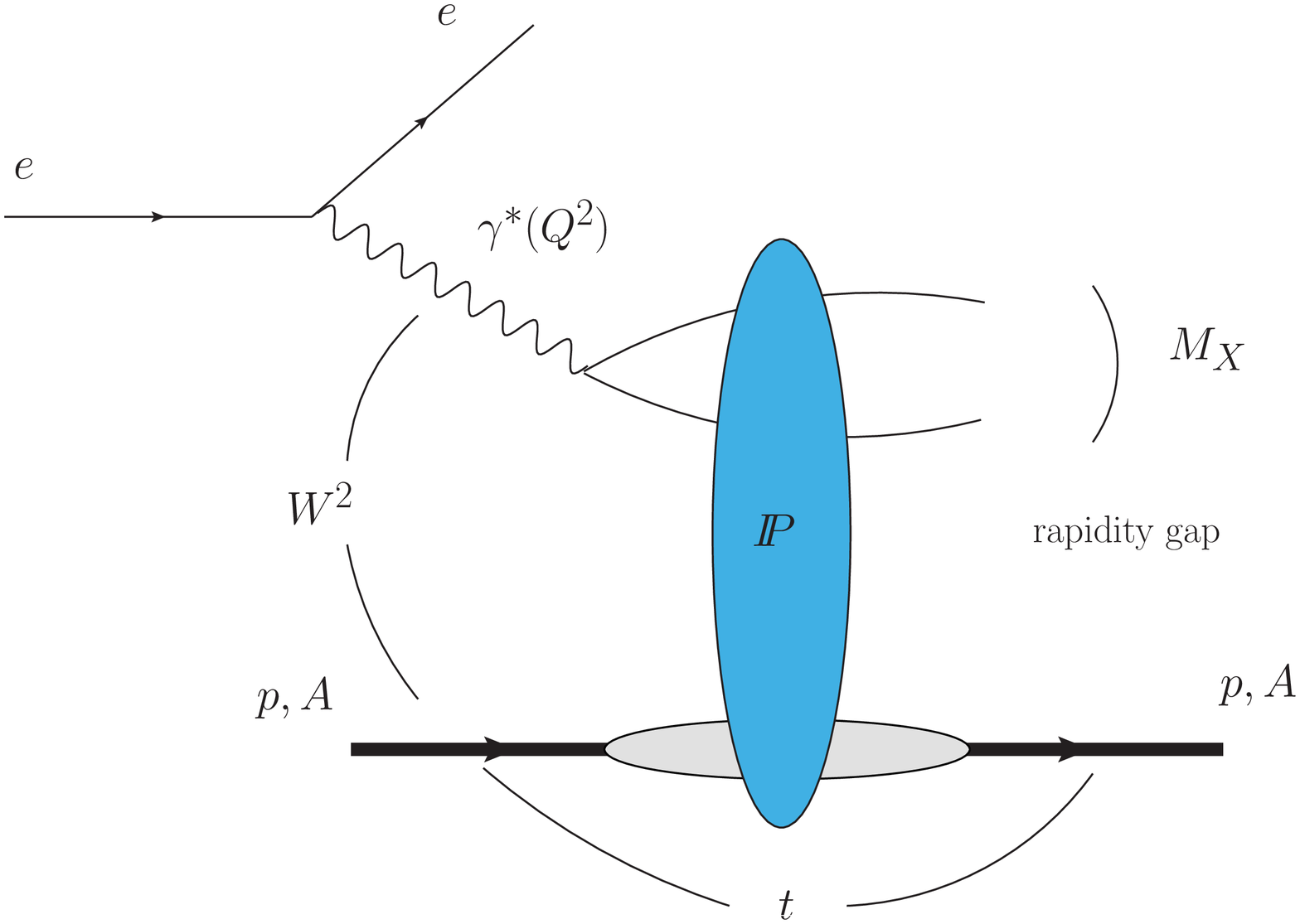}} & 
{\includegraphics[width=0.5\textwidth]{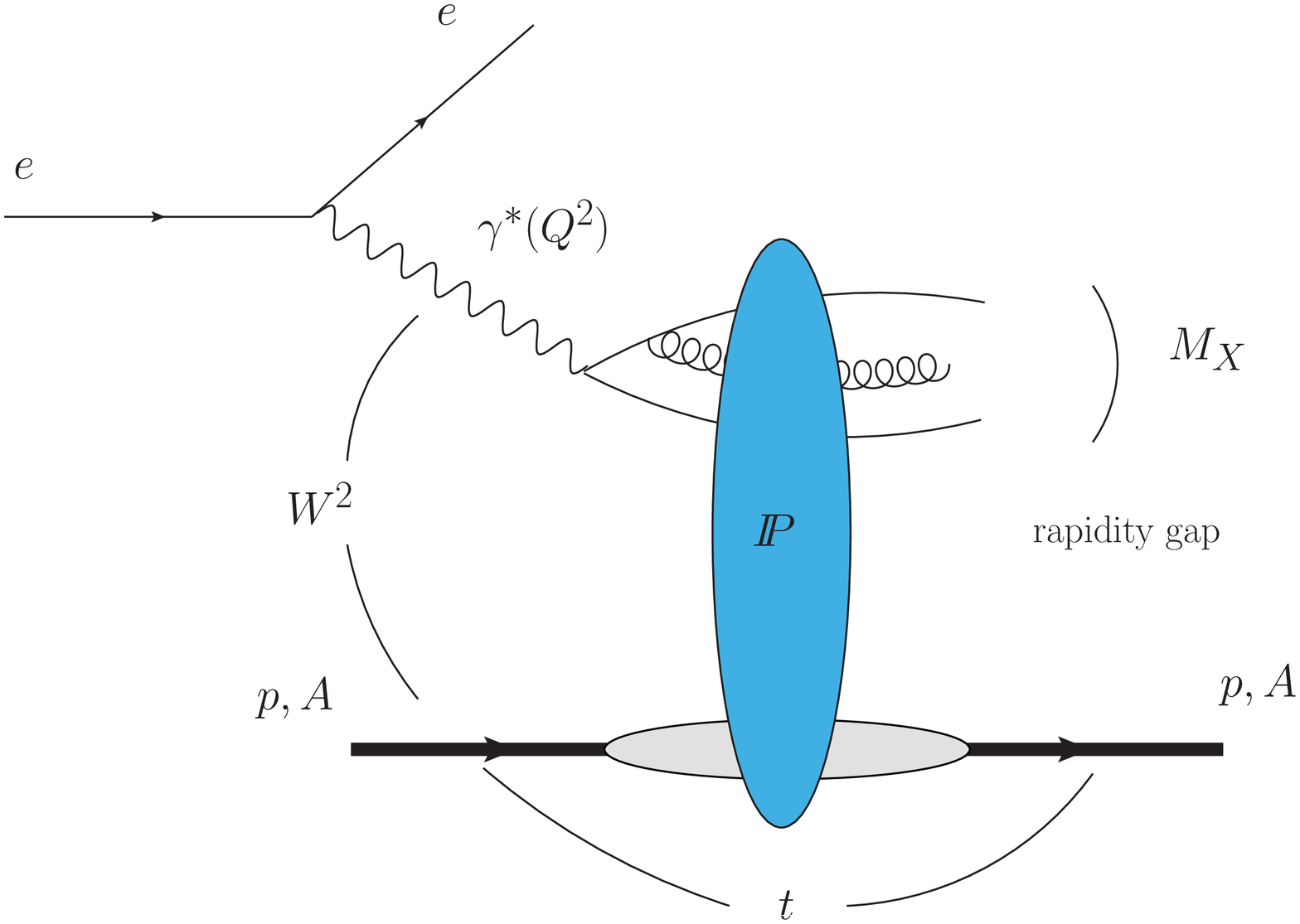}} \\ 
 (a) & (b)  
\end{tabular}                                                                                                                    
\caption{Diffractive deeply inelastic electron-hadron scattering in the color-dipole formalism, which assumes that the virtual photon fluctuates into a colorless parton Fock state, which interacts elastically with the proton or nucleus. The contributions of $q\bar{q}$ and $q\bar{q}g$ parton Fock states are presented in the left and right panels, respectively.}
\label{fig:dipole}
\end{figure}

Currently, there have been many attempts to describe the diffractive part of the deeply inelastic cross section within the pQCD, see e.g. Refs.~\cite{GBW,fss,MRW,Brod,Frankfurt:2011cs,Armesto:2019gxy}. One of the most successful approaches is the saturation one~\cite{GBW,Marquet} based on the dipole picture of DIS~\cite{dipole,dipole2}. It naturally incorporates the description of both inclusive and diffractive events in a common theoretical framework, as the same dipole-target scattering amplitude enters in the formulation of the inclusive and diffractive cross sections. In this formalism, the total and diffractive cross sections are given by
\begin{eqnarray}
\sigma_{\rm tot}(x,Q^2) & = & \sum_{i = L,\,T} \int \mathrm{d}z \, \mathrm{d}^2r |\Psi_{i}^{\gamma{*}}(z,r)|^{2}  \int \mathrm{d}^2\rb_t\frac{\mathrm{d}\sigma}{\mathrm{d}^2\rb_t} \,,\\
\sigma_{\rm diff}(x,Q^2) & = &  \frac{1}{4} \sum_{i = L,\,T} \int \mathrm{d}z \, \mathrm{d}^2r |\Psi_{i}^{\gamma{*}}(z,r)|^{2} \int \mathrm{d}^2\rb_t \left(\frac{\mathrm{d}\sigma}{\mathrm{d}^2\rb_t}\right)^2 \, ,
\end{eqnarray}
where the functions $|\Psi^{\gamma{*}}_{T,L}(r,z)|^2$ represent the probability of the photon with transverse (T) or longitudinal (L) polarization to split into a $q\bar{q}$ pair; these functions can be calculated perturbatively and are expressed by
\begin{eqnarray}
|\Psi_{T}^{\gamma{*}}(z,r)|^{2} & = & \frac{6\alpha_{em}}{(2\pi)^{2}}\sum_{f}e^{2}_{f}\left\{[z^{2}+(1-z)^{2}]\epsilon^{2}K_{1}^{2}(\epsilon{r})+m_{f}^{2}K^{2}_{0}(\epsilon{r})\right\} \,, \\
|\Psi_{L}^{\gamma{*}}(z,r)|^{2} & = & \frac{6\alpha_{em}}{(2\pi)^{2}}\sum_{f}4e^{2}_{f}Q^{2}z^{2}(1-z)^{2}K^{2}_{0}(\epsilon{r}) \,,
\end{eqnarray}
where $r$ is the size of the $q\bar{q}$, $z$ and $(1-z)$ are the momentum fractions of the original photon momentum carried by the quark and anti-quark, respectively, and $m_{f}$ and $e_{f}$ are the mass and the charge of a quark with flavor $f$. Moreover, ${\mathrm{d}\sigma}/{\mathrm{d}^2\rb_t}$ denotes the dipole cross section for its scattering off the target at an impact parameter $\rb_t$; this cross section is related to the dipole-target scattering amplitude ${\cal{N}}(x,\rr,\rb_t)$ by
\begin{eqnarray}
\frac{\mathrm{d}\sigma}{\mathrm{d}^2\rb_t} = 2 \,{\cal{N}}(x,\rr,\rb_t) \,.
\end{eqnarray}

In addition, it is possible to extend the color-dipole formalism to estimate the diffractive structure functions. For that, one needs to compute the reduced diffractive cross section $\sigma^{D(3)}_{r}(x_{\pom},\beta,Q^{2})$. In Ref.~\cite{GBW}, the authors have derived expressions for $F_2^{D (3)}$ directly in the transverse momentum space and then transformed them to impact-parameter space where the dipole approach can be applied. Following Refs.~\cite{GBW,Marquet,Kowalski_prc}, we assume that the virtual photon fluctuates into a colorless parton Fock state, which interacts elastically with the proton or the nucleus. We include the contributions of $q\bar{q}$ and $q\bar{q}g$ states, with the associated diffractive processes being represented in Fig.~\ref{fig:dipole}.  Both processes are characterized by the presence of a rapidity gap in the final state due to the color singlet exchange. As a consequence, the diffractive structure function can be expressed by
\begin{equation}
F_2^{D (3)} (Q^{2}, \beta, x_{I\!\!P}) = F^{D}_{q\bar{q},L} + F^{D}_{q\bar{q},T} + F^{D}_{q\bar{q}g,T},
\label{soma}
\end{equation}
where $T$ and $L$ refer to the polarization of the virtual photon. For the $q\bar{q}g$ contribution, only the transverse polarization is considered, since the longitudinal counterpart has no leading logarithm in $Q^2$. The computation of the different contributions was done in Refs.~\cite{GBW, Marquet, Kowalski_prc} and here we only quote the final results. The transverse component of the $q\bar{q}$ contribution is given by
\begin{equation}
x_{\pom}F_{q\bar{q},T}^{D}(x_{\pom},\beta,Q^{2})=\frac{N_{c}Q^{4}}{16\pi^{3}\beta}\sum_{f}{e_{f}^{2}\int_{z_{0}}^{1/2}\mathrm{d}zz(1-z)\Big\{\epsilon^{2}[z^{2}+(1-z)^{2}]\Phi_{1}+m_{f}^{2}\Phi_{0}\Big\}} \,,
\end{equation}
while the longitudinal contribution is
\begin{equation}
x_{\pom}F_{q\bar{q},L}^{D}(x_{\pom},\beta,Q^{2})=\frac{N_{c}Q^{6}}{4\pi^{3}\beta}\sum_{f}{e_{f}^{2}\int_{z_{0}}^{1/2}\mathrm{d}zz^{3}(1-z)^{3}\Phi_{0}} \,,
\end{equation}
where $\epsilon^{2}=z(1-z)Q^{2}+m_{f}^{2}$ and the variable $z_0$ is defined by
\begin{equation}
z_{0}=\frac{1}{2}\Bigg(1-\sqrt{1-\frac{4m_{f}^{2}}{M_{X}^{2}}}\Bigg) \,.
\end{equation}
Moreover, the auxiliary functions $\Phi_{0,1}$ are expressed as follows
\begin{equation}
\Phi_{0,1}=\int{\mathrm{d}^{2}\rb_{t}\Bigg[\int_{0}^{\infty}\mathrm{d}rrK_{0,1}(\epsilon r)J_{0,1}(k r)\frac{\mathrm{d}\sigma}{\mathrm{d}^{2}\rb_{t}}({\rb_{t}},{\rr},x_{\pom})\Bigg]^{2}} \,,
\end{equation}
with $k^2 = z(1 - z) M_X^2 - m_f^2$.\\
Finally, the transverse $q\bar{q}g$ component is given by
\begin{eqnarray}
\hspace{-0.5cm}x_{\pom}F_{q\bar{q}g,T}^{D}(x_{\pom},\beta,Q^{2})=\frac{\alpha_{s}\beta}{8\pi^{4}}\sum_{f}{e_{f}^{2}}\int \mathrm{d}^{2}\rb_{t} \int_{0}^{Q^{2}}\mathrm{d}\kappa^{2} \int_{\beta}^{1}\mathrm{d}z\left\{\kappa^{4}\ln{\frac{Q^{2}}{\kappa^{2}}}\Bigg[\bigg(1-\frac{\beta}{z}\bigg)^{2}+\bigg(\frac{\beta}{z}\bigg)^{2}\Bigg]\right.\nonumber \\
\left. \Bigg[\int_{0}^{\infty}\mathrm{d}rr \frac{\mathrm{d}\sigma_g}{\mathrm{d}^{2}\rb_t}K_{2}(\sqrt{z}\kappa {r})J_{2}(\sqrt{1-z}\kappa {r})\Bigg]^{2}\right\}\, ,
\label{qqg} 
\end{eqnarray}
where
\begin{equation}
\frac{\mathrm{d}\sigma_g}{\mathrm{d}^{2}\rb_{t}}=2\left[1-\Big(1-\frac{1}{2}\frac{\mathrm{d}\sigma}{\mathrm{d}^{2}\rb_t}\Big)^{2}\right]\,. 
\end{equation}

As pointed out in Ref.~\cite{Marquet}, at small $\beta$ and low $Q^2$, the leading $\ln (1/\beta)$ terms should be resumed and the above expression should be modified. However, as a description with the same quality using  Eq.~(\ref{qqg}) is possible by adjusting the coupling \cite{Marquet} and our focus will be in the predictions for medium values of $\beta$, in what follows we will use this expression for our studies.

In the color-dipole formalism, the energy, photon virtuality and atomic number dependencies of diffractive observables are fully determined by the evolution of $\cal{N}$ and, consequently, strongly dependent on the description of the QCD dynamics at small-$x$ and large-$A$. As discussed in the introduction, the Balitsky--Kovchegov (BK) equation is a nonlinear evolution equation in rapidity $Y$ for the dipole-hadron scattering amplitude and is given by~\cite{BAL,kov}
\begin{equation}\label{eq:bklo}
\frac{\partial {\cal{N}}(\rr,\rb_t,Y)}{\partial Y} = \int {\rm d}\bm{\rr_1}\, K(\bm{\rr,\rr_1,\rr_2}) [{\cal{N}}(\rr_1, \rb_1, Y)+{\cal{N}}(\rr_2, \rb_2, Y)-{\cal{N}}(\rr, \rb_t, Y)-{\cal{N}}(\rr_1, \rb_1, Y){\cal{N}}(\rr_2, \rb_2, Y)].
\end{equation}
In the following, $r \equiv |\rr|$, $r_1 \equiv |\rr_1|$ and $r_2 \equiv |\rr_2|$ are the transverse sizes of the original dipole and of the two daughter dipoles, respectively, and the $b_i \equiv \rb_i$ are the corresponding impact-parameter vectors. During the last years, different functional forms have been proposed for the kernel $K(\bm{\rr,\rr_1,\rr_2})$ of the BK equation, considering e.g., the corrections that take into account the running of the coupling constant as well as the resummation of collinear logarithms. The studies performed in Refs.~\cite{soyez_coll,alba_coll}, which disregarded the dependence on the impact parameter, demonstrated that it is possible to obtain a good description of the inclusive HERA data by taking into account these corrections to the kernel. However, in order to describe in detail exclusive processes, it is fundamental to take into account also the impact-parameter dependence of the dipole-target scattering amplitude. In Refs.~\cite{cepila_nucleon,cepila_nucleon1}, the BK equation was solved for a proton target including the dependence on impact parameter and using the collinearly-improved kernel. The authors have demonstrated that the contribution coming from the large impact parameters is strongly suppressed by the collinear corrections and that the HERA data for the $F_2$ structure function and for the exclusive vector meson production are reasonably well described. More recently, this approach was extended for nuclear targets in Refs.~\cite{cepila_nuclear,cepila_nuclear2}. In what follows, we  discuss the main characteristics of these solutions, which are used as input in the calculation of the diffractive observables.

\begin{figure}[t]
 {\includegraphics[width=0.7\textwidth]{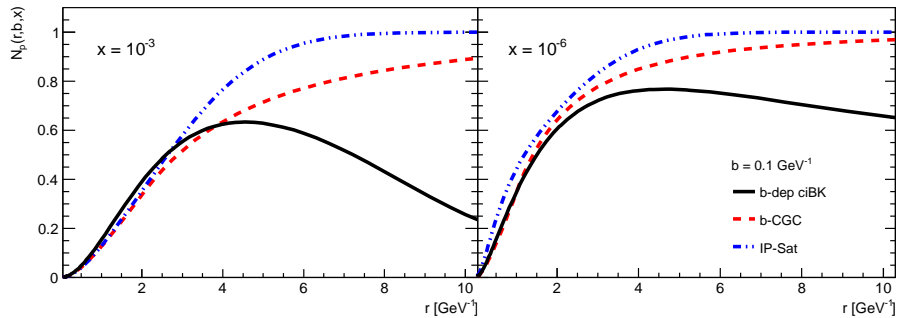}}
\caption{Dipole-proton scattering amplitude for a fixed impact parameter ($b = 0.1$ GeV$^{-1}$) and two values of $x$. }
\label{fig:nproton}
\end{figure}

In Fig.~\ref{fig:nproton}, we present the solutions of the impact-parameter dependent BK equation with the collinearly-improved kernel for a proton target, denoted as b-dep ciBK hereafter, for a fixed value of $b$ and two values of $x$. As in Refs.~\cite{cepila_nucleon,cepila_nucleon1}, the initial condition is given by a combination of the GBW model~\cite{GBW} for the dependence on the dipole size $r$ and a Gaussian distribution for the impact-parameter dependence. The parameters have been fixed using HERA data for $F_2$ and for the $|t|$-distribution of the $J/\Psi$ photoproduction. For comparison, we also present the predictions from the IP-Sat and b-CGC models (see e.g. Refs.~\cite{kmw,ipsat4,amir}), which are phenomenological models based on the CGC physics and that are also able to describe HERA data. Although the predictions give similar results for small dipoles, they strongly differ at large-$r$. Such difference is smaller for smaller values of the Bjorken-$x$, with the onset of saturation being slower in the b-dep ciBK case. This result indicates that observables sensitive to large dipole sizes will be sensitive to the modeling of $\cal{N}$.

\begin{figure}[t]
{\includegraphics[width=0.7\textwidth]{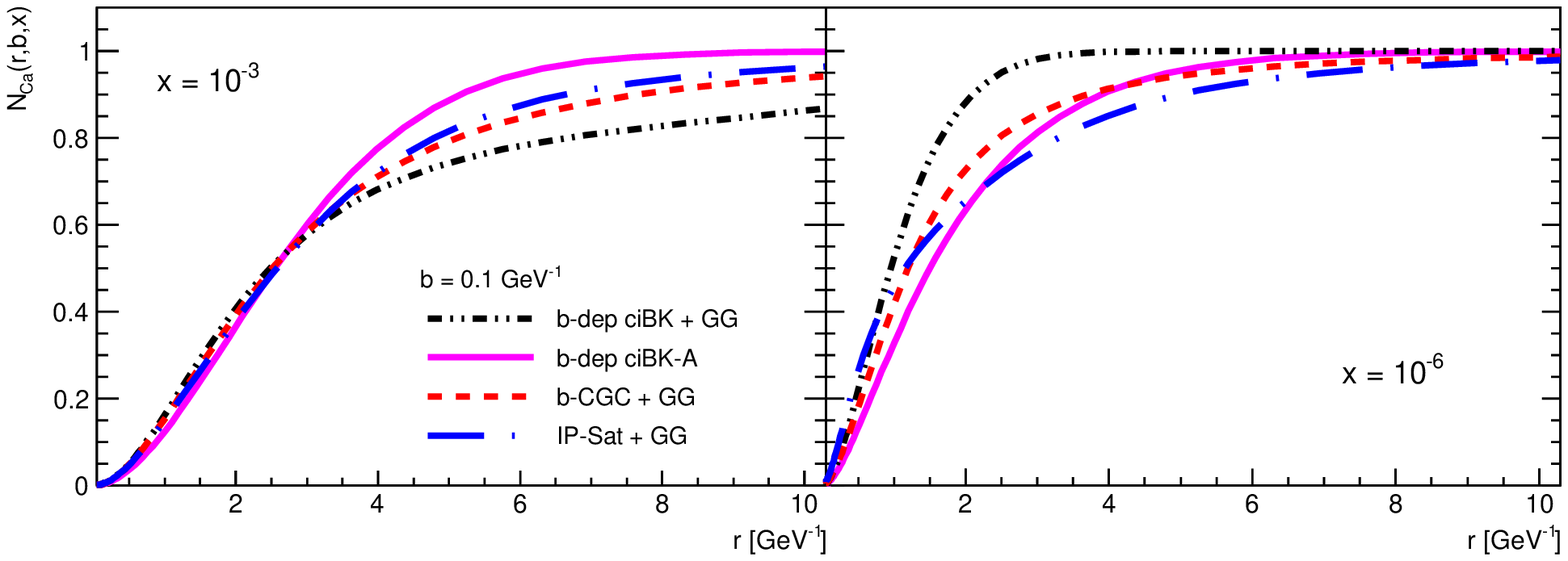}} \\ 
{\includegraphics[width=0.7\textwidth]{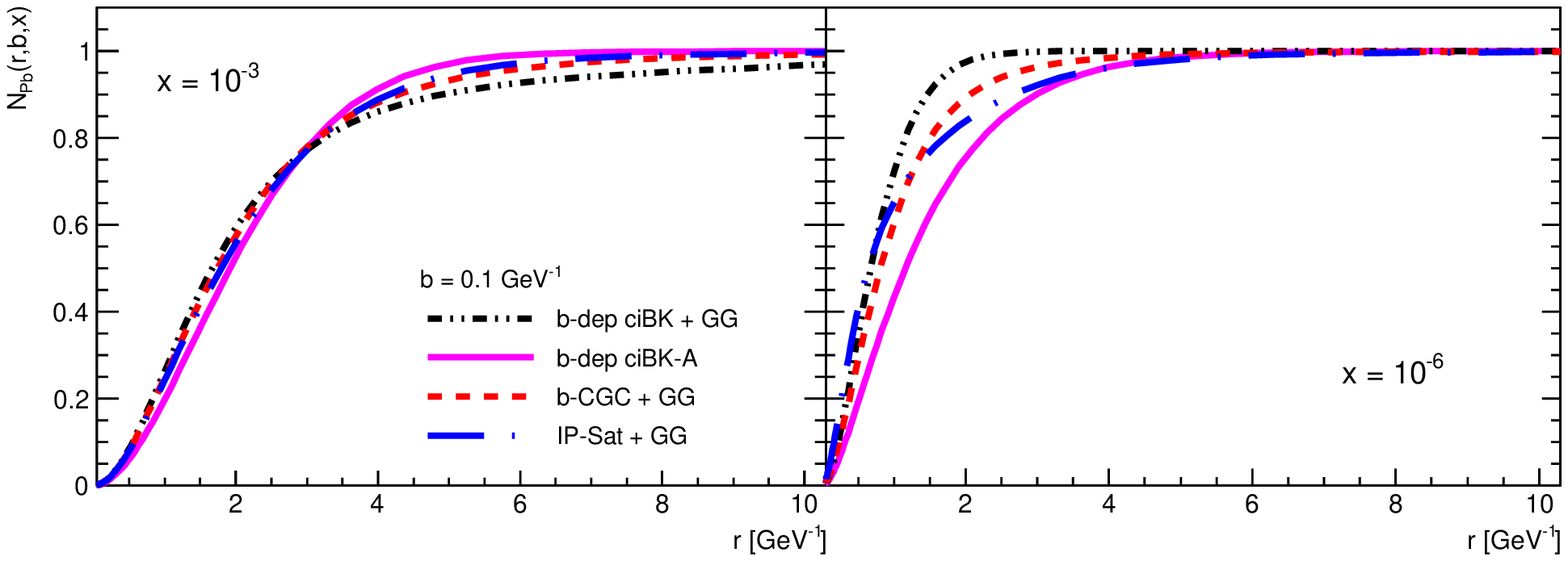}} 
\caption{Dipole-nucleus scattering amplitude for a fixed impact parameter $b = 0.1$ GeV$^{-1}$ and two values of $x$. Predictions for $A = 40 \,(208)$ are presented in the upper (lower) panels.}
\label{fig:nnucleus}
\end{figure}

In Fig.~\ref{fig:nnucleus}, the BK solution for a nuclear target, denoted b-dep ciBK-A hereafter, is presented for a fixed value of $b$ and two values of $x$. As in Ref.~\cite{cepila_nuclear}, the initial condition is  given by
\begin{eqnarray}
{\cal{N}}_A(\rr,\rb_A,Y=0) = 1 - \exp \left[-  \frac{1}{2} \frac{Q_{s0}^2(A)}{4} r^2 T_A(\rb_{q_1},\rb_{q_2}) \right] \,,
\end{eqnarray}
where $\rb_A$ is the dipole-nucleus impact parameter, $Y = \ln (x_0/x)$ with $x_0 = 0.008$, $\rb_{q_i}$ are the impact parameters with respect to the dipole constituents, and $Q_{s0}^2$ is a free parameter determined for each value of $A$ by the comparison between the dipole predictions for $F_2^A$ and those obtained using the collinear formalism and the EPPS16 parametrization~\cite{epps} for $Y = 0$. The nuclear profile $T_A(\rb_{q_1},\rb_{q_2})$ is given by
\begin{eqnarray}
T_A(\rb_{q_1},\rb_{q_2}) = k \left[T_A(\rb_{q_1}) + T_A(\rb_{q_2}) \right],
\end{eqnarray}
where the individual profiles $T_A(\rb_{q_i})$
are described by a Woods-Saxon distribution and $k$ is the factor which ensures $kT_A(0) = 1$. For comparison, we also present the predictions for ${\cal{N}}_A$ derived using the Glauber-Gribov (GG) formalism~\cite{glauber,gribov,mueller,Armesto:2002ny}, which implies that ${\cal{N}}_A(\rr,\rb_A,Y)$ is given by
\begin{eqnarray}
{\cal{N}}_A(\rr,\rb_A,Y) =  1 - \exp \left[-\frac{1}{2}  \, \sigma_{dp}(Y,\rr^2) \,T_A(\rb_A)\right] \,,
\label{enenuc}
\end{eqnarray}
where $\sigma_{dp}$ is the dipole-proton cross section, which is expressed in terms of the dipole-proton scattering amplitude as follows
\begin{eqnarray}
\sigma_{dp}(Y,\rr^2) = 2 \int \mathrm{d}^2\rb_p \, {\cal{N}}_p(\rr,\rb_p,Y) \,,
\end{eqnarray}
where $\rb_p$ is the impact parameter for the dipole-proton interaction. In order to estimate the dependence of our predictions on the treatment of the nonlinear effects, we compare the solution of the BK equation for the nuclear case with those derived using the BK solution for the proton as the input in Eq.~(\ref{enenuc}), the said predictions being denoted as b-dep ciBK + GG in what follows. In addition, we also  present the predictions obtained using the IP-Sat and b-CGC models as input in the GG formula, which are denoted as IP-Sat + GG and b-CGC + GG, respectively. The results presented in Fig.~\ref{fig:nnucleus} for $A = 40$ (upper panels) and $A = 208$ (lower panels) indicate that the onset of the nonlinear (saturation) effects (${\cal{N}} \approx 1$) occurs at smaller values of $r$ for heavier nuclei and for smaller values of $x$. Such result is indeed not surprising, since the nuclear saturation scale is expected to increase with the atomic number of the nucleus and with the increasing rapidity $Y$. The BK equation for the nuclear case predicts a faster onset of total saturation for $x = 10^{-3}$ than the predictions derived using the GG formula. On the other hand, for $x = 10^{-6}$, the b-dep ciBK + GG prediction saturates at smaller values of $r$, with the transition between the linear (small-$r$) and nonlinear (large-$r$) regimes being strongly model dependent. Such result motivates the analysis of the impact of these distinct descriptions for the nonlinear effects on the diffractive observables that can be measured in future $eA$ collisions.

\section{Results}
\label{sec:res}  

\begin{figure}[t]
\begin{tabular}{ccc}
 {\includegraphics[width=0.33\textwidth]{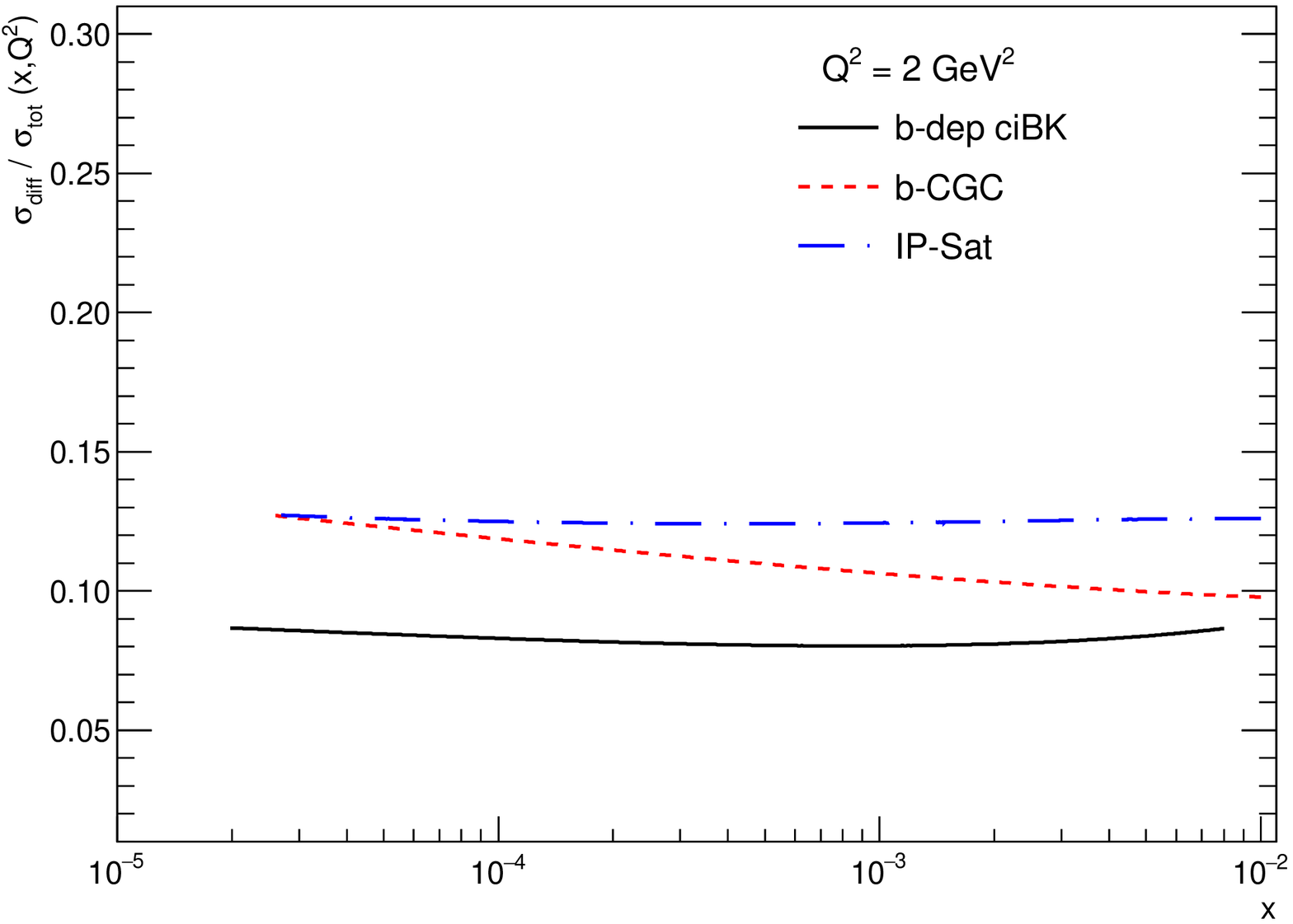}} & 
{\includegraphics[width=0.33\textwidth]{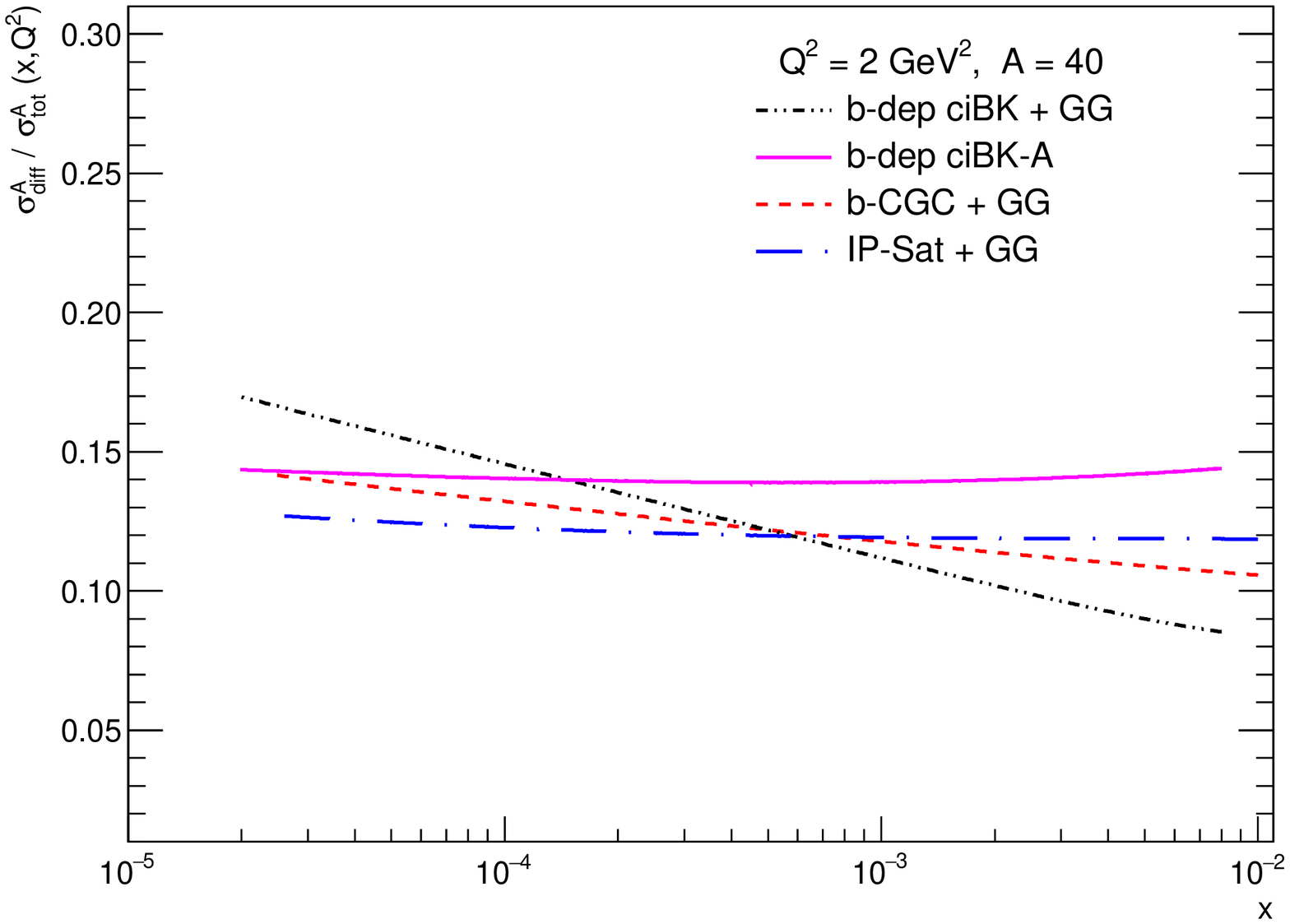}} &
{\includegraphics[width=0.33\textwidth]{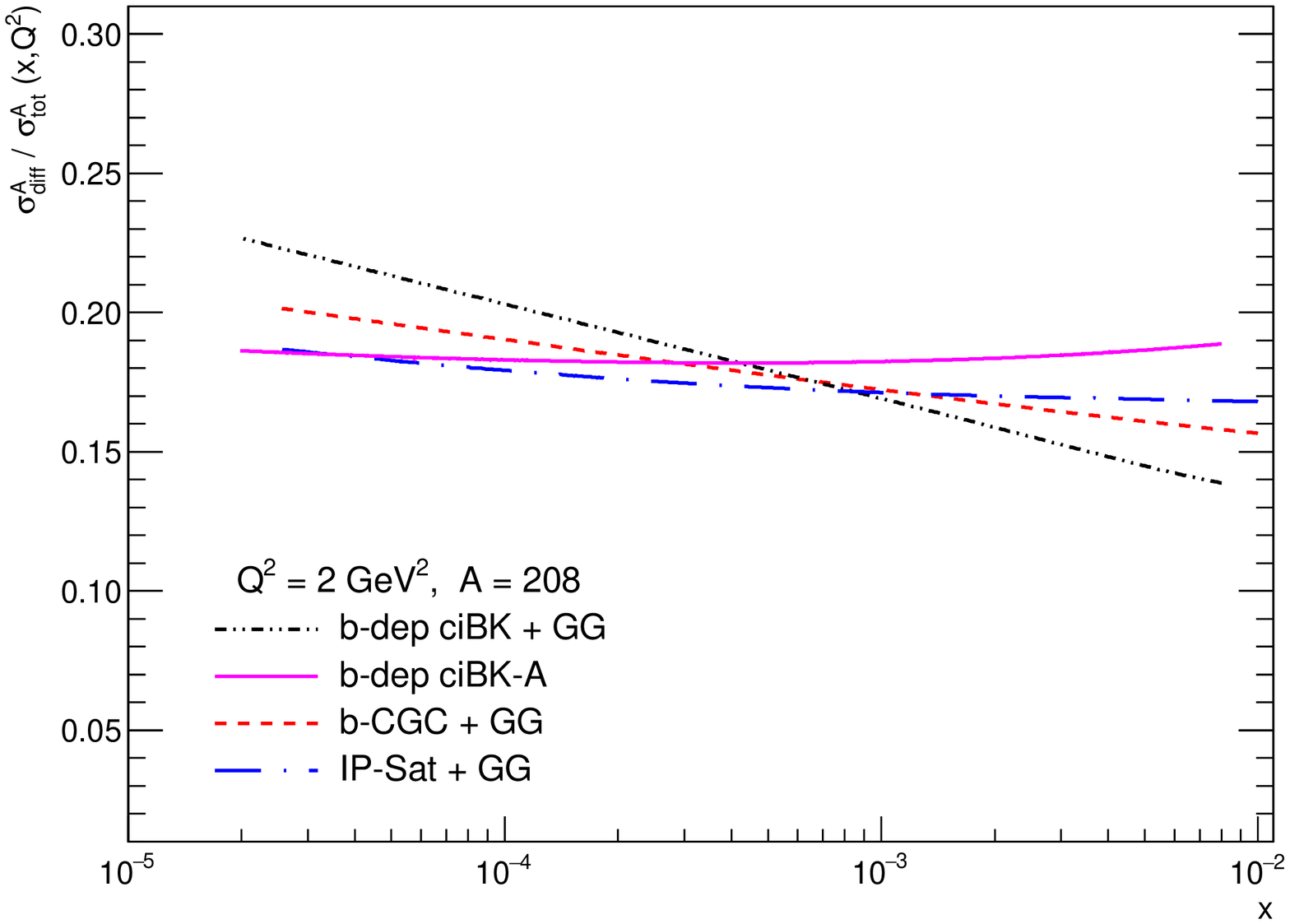}} \\ 
 (a) & (b)  & (c)
\end{tabular}

\caption{Predictions for the $x$ dependence of the ratio $R_{\sigma} = \sigma_{\rm diff}/\sigma_{\rm tot}$ for (a) $A = 1$, (b) $A = 40$ and (c) $A = 208$, considering distinct models for the dipole-target scattering amplitude and a fixed photon virtuality ($Q^2 = 2$ GeV$^2$).}
\label{fig:ratio}
\end{figure}

Initially, let's investigate the impact of the nonlinear effects on the ratio between the diffractive and total cross sections, defined by $R_{\sigma} = \sigma_{\rm diff}/\sigma_{\rm tot}$, which allows us to analyze if the diffractive and inclusive processes have a similar   energy-, $x$-, $Q^2$- and $A$-dependencies. 
The experimental results from HERA indicate a very similar energy-dependence of the diffractive and the total cross section. It is important to emphasize that  saturation physics provides a simple explanation for this finding. As   shown e.g. in Refs.  \cite{GBW,simone2}, the saturation effects suppress the  contribution of the large-size dipoles, associated to nonperturbative physics, and implies that the energy dependence is similar for inclusive and diffractive processes. In contrast, to explain this aspect of data using a description based on the collinear factorization is non-trivial. In this case the energy dependence of the inclusive and diffractive cross sections is controlled by the $x$-dependence of the ordinary and the diffractive parton densities, which is not predicted by the theory. Another important aspect is that in the black-disc limit, where $\frac{\mathrm{d}^2\sigma}{\mathrm{d}^2\rb} \rightarrow 2$, we have that $R_{\sigma} \rightarrow 1/2$. Therefore, this observable can be considered as a measure of how close to the black-disc limit we are. Previous calculations have demonstrated that $R_{\sigma}$ increases with the atomic number $A$~\cite{Nikolaev,simone2,Nik_schafer,Kowalski_prc}. The results presented in Fig.~\ref{fig:ratio} agree with these expectations, with the distinct models predicting that $\approx 20 \%$ of the events will be diffractive in $ePb$ collisions and will have a $x$-dependence similar to the inclusive one. We have verified that this value decreases with the increase of the photon virtuality. For the proton case, the b-dep ciBK model predicts the smaller amount of diffractive events. On the other hand, for the nuclear case, the b-dep ciBK-A and b-dep ciBK + GG results differ in their predictions for the $x$-dependence of the ratio, with the b-dep ciBK-A one being almost flat, while the b-dep ciBK + GG model predicts that $R_{\sigma}$ increases at smaller values of $x$.

 \begin{figure}[t]
\begin{tabular}{ccc}
 {\includegraphics[width=0.33\textwidth]{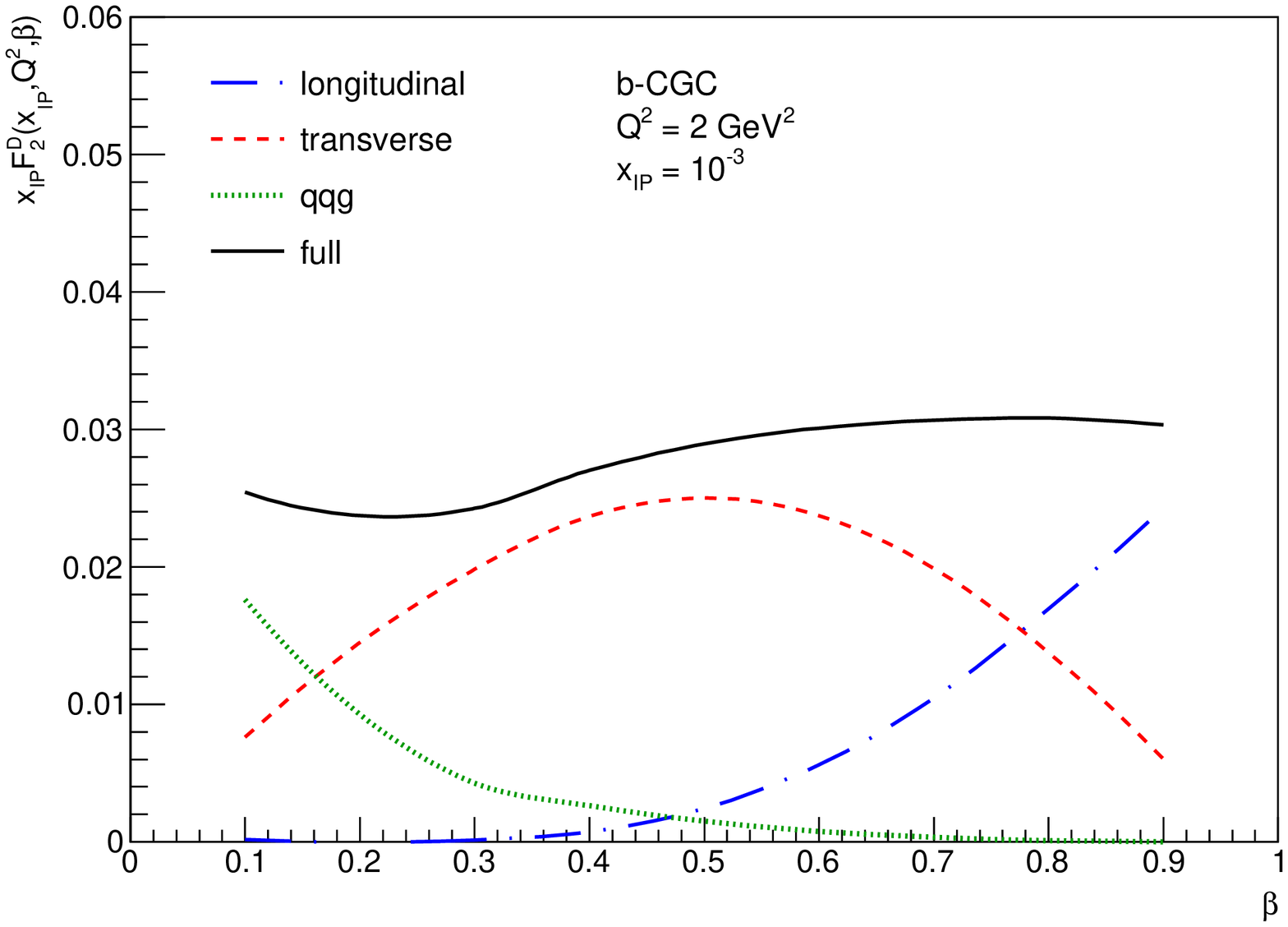}} & 
{\includegraphics[width=0.33\textwidth]{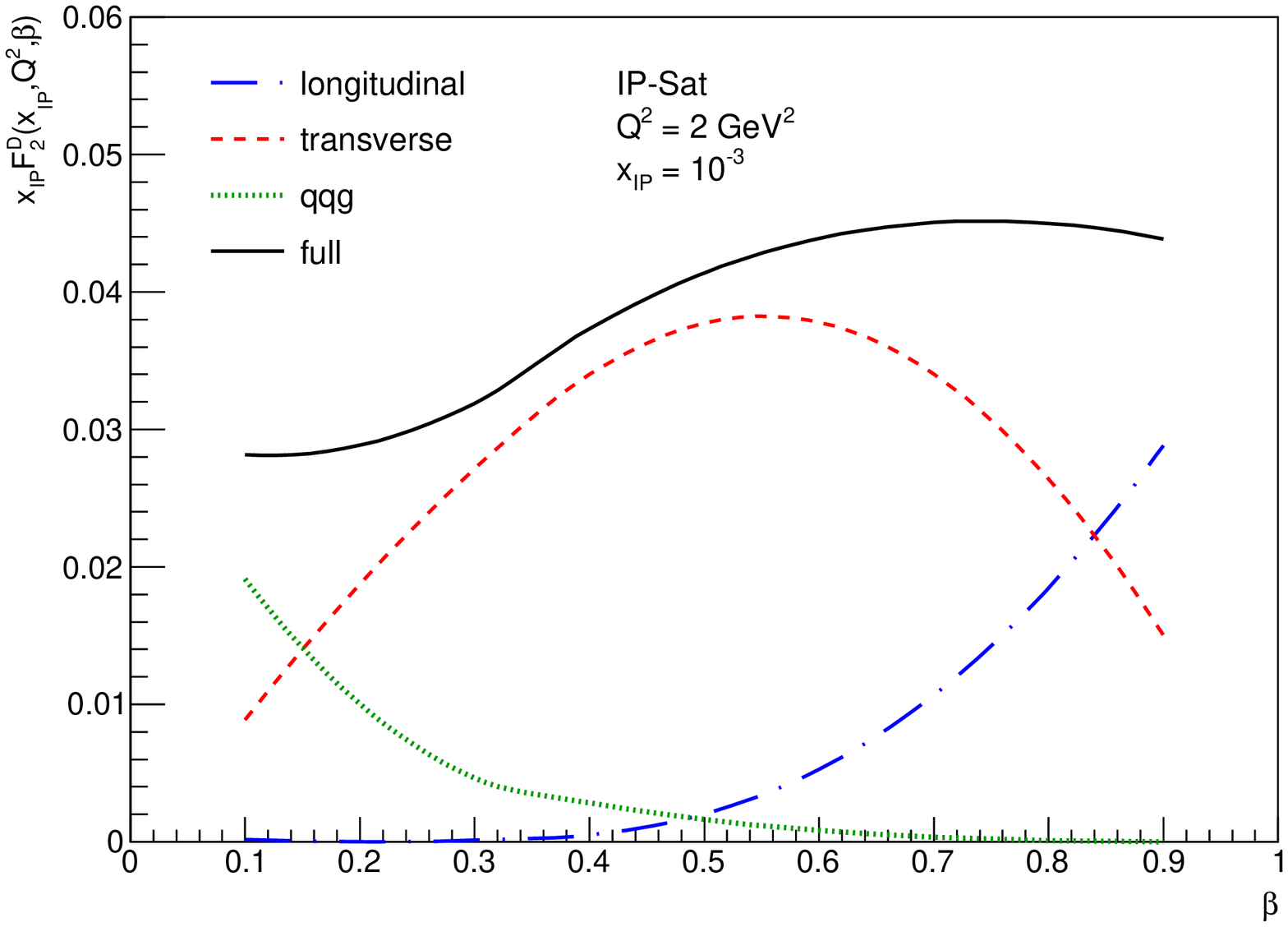}} &
{\includegraphics[width=0.33\textwidth]{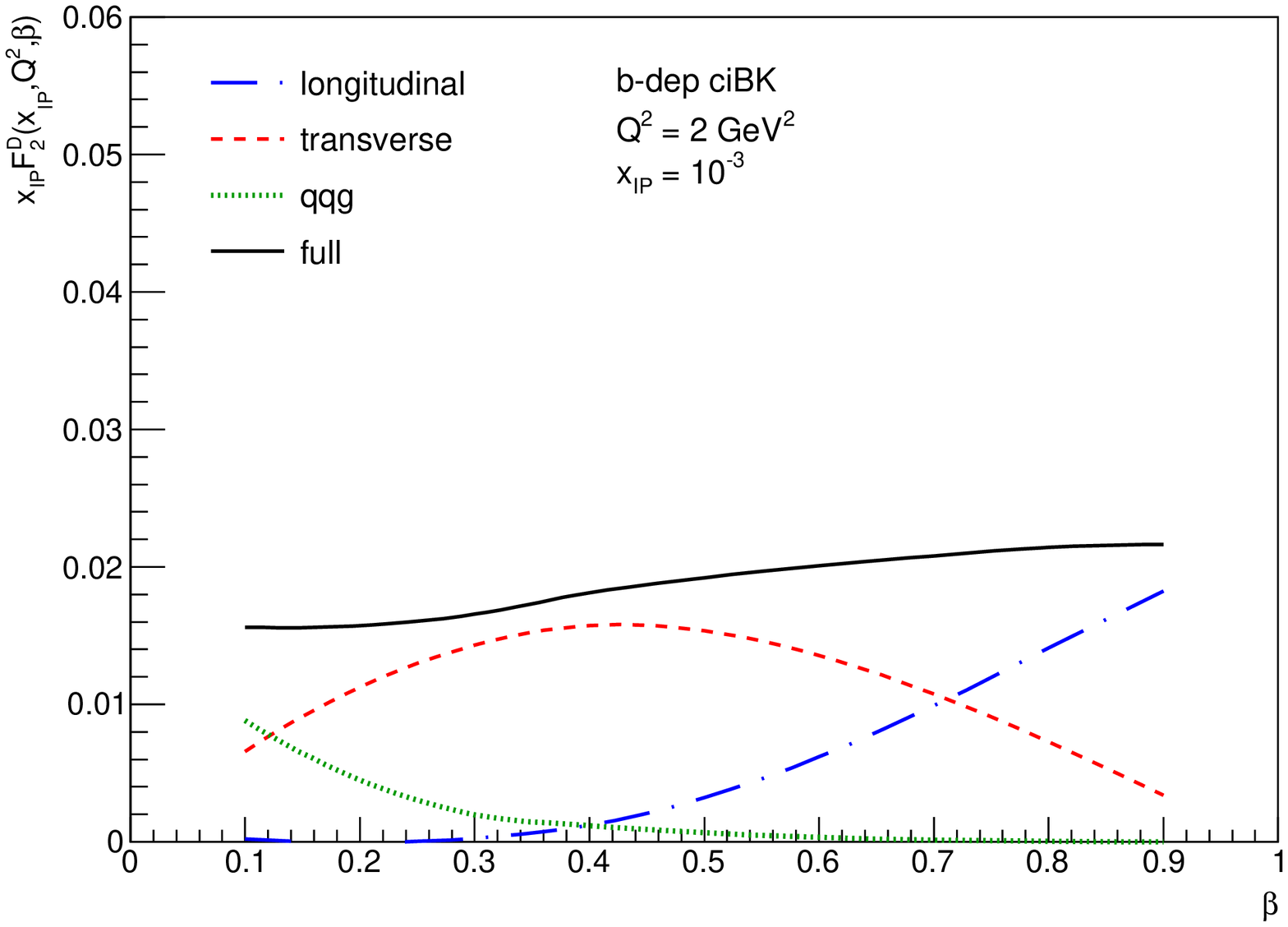}} \\
\end{tabular}                                                                                                                       
\caption{Predictions for the $\beta$  dependence of the different components of $F_2^{D (3)}$ considering distinct models for the dipole-proton scattering amplitude.}
\label{fig:f2d3_comp}
\end{figure}

The diffractive cross section $ep \rightarrow eXY$ has been measured by the H1 and ZEUS experiments at HERA by tagging the proton in the final state ($Y = p$) or by selecting events with a large rapidity gap between the systems $X$ and $Y$. In our study, we will focus on the case where the proton or the nucleus remains intact in the final state. Moreover, our analysis will focus on the kinematic range of small $Q^2$ and $x_{I\!\!P} \le 10^{-2}$, which is the range expected to be probed in the future EICs and where the nonlinear effects are predicted to significantly contribute. In Fig.~\ref{fig:f2d3_comp}, we present our predictions for the $\beta$ dependence of the diffractive structure function $F_2^{D (3)}$ at fixed $Q^2$ and $x_{\pom}$, considering different models for the dipole-proton scattering amplitude. Although the general structure of the $\beta$-spectrum for the distinct components is determined by the photon wave function, we have that the normalization of these distinct terms depends on the modeling of the dipole-proton interaction, with the b-dep ciBK prediction being smaller than the other two phenomenological models. Moreover, this model predicts a smaller contribution of the $q\bar{q}g$ component at small-$\beta$. 

\begin{figure}[t]
\begin{tabular}{ccc}
 {\includegraphics[width=0.33\textwidth]{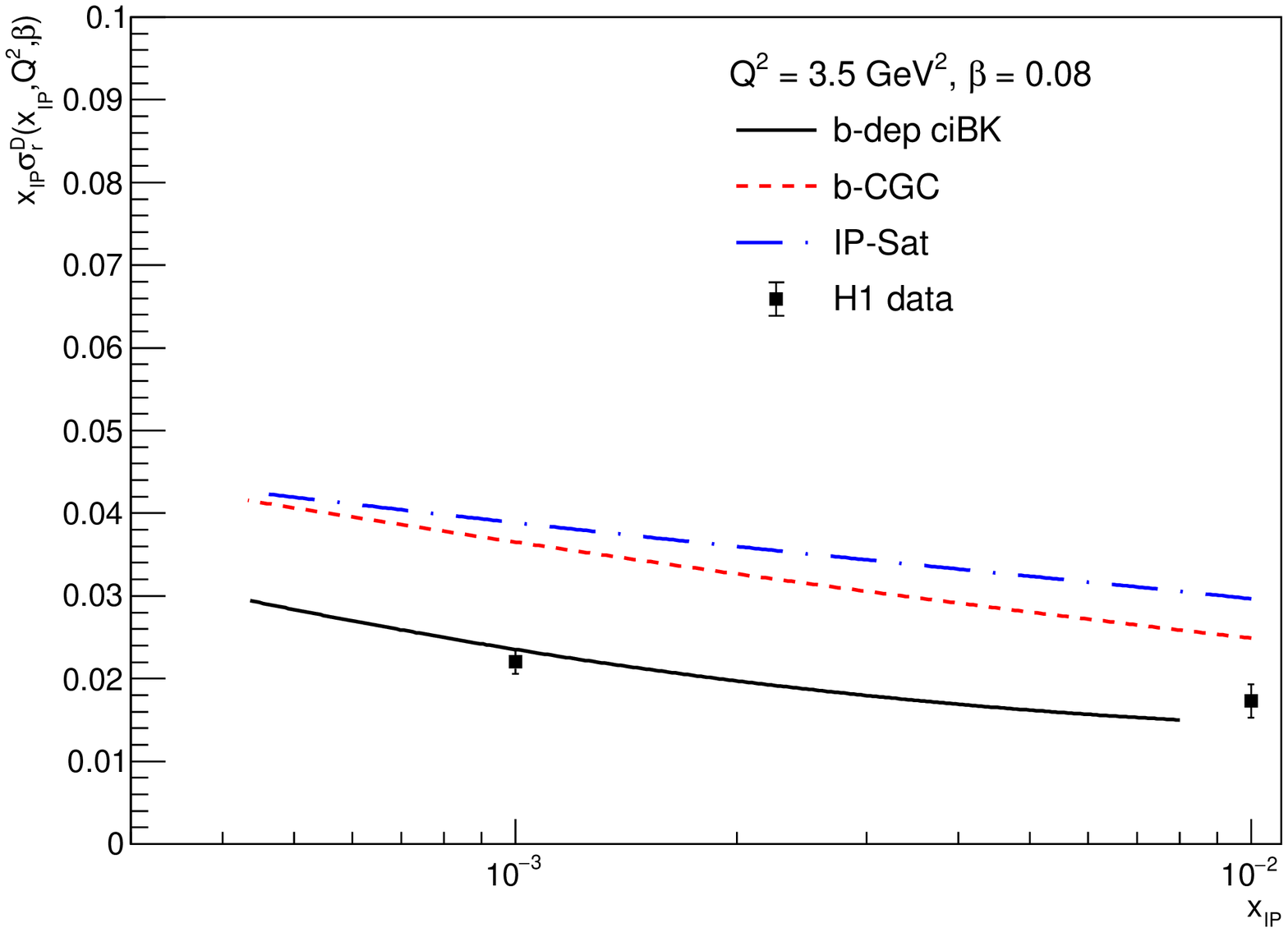}} & 
{\includegraphics[width=0.33\textwidth]{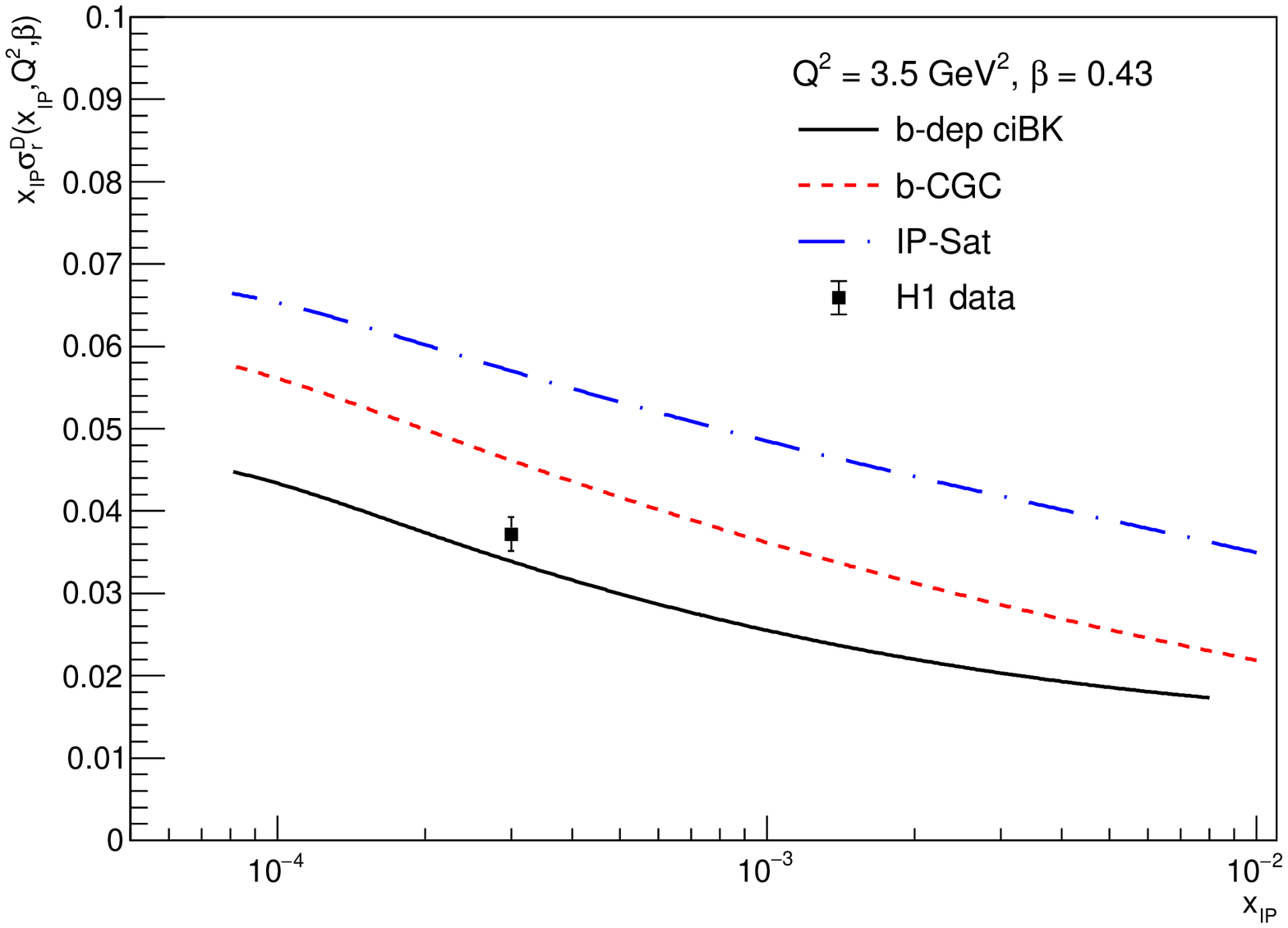}} &
{\includegraphics[width=0.33\textwidth]{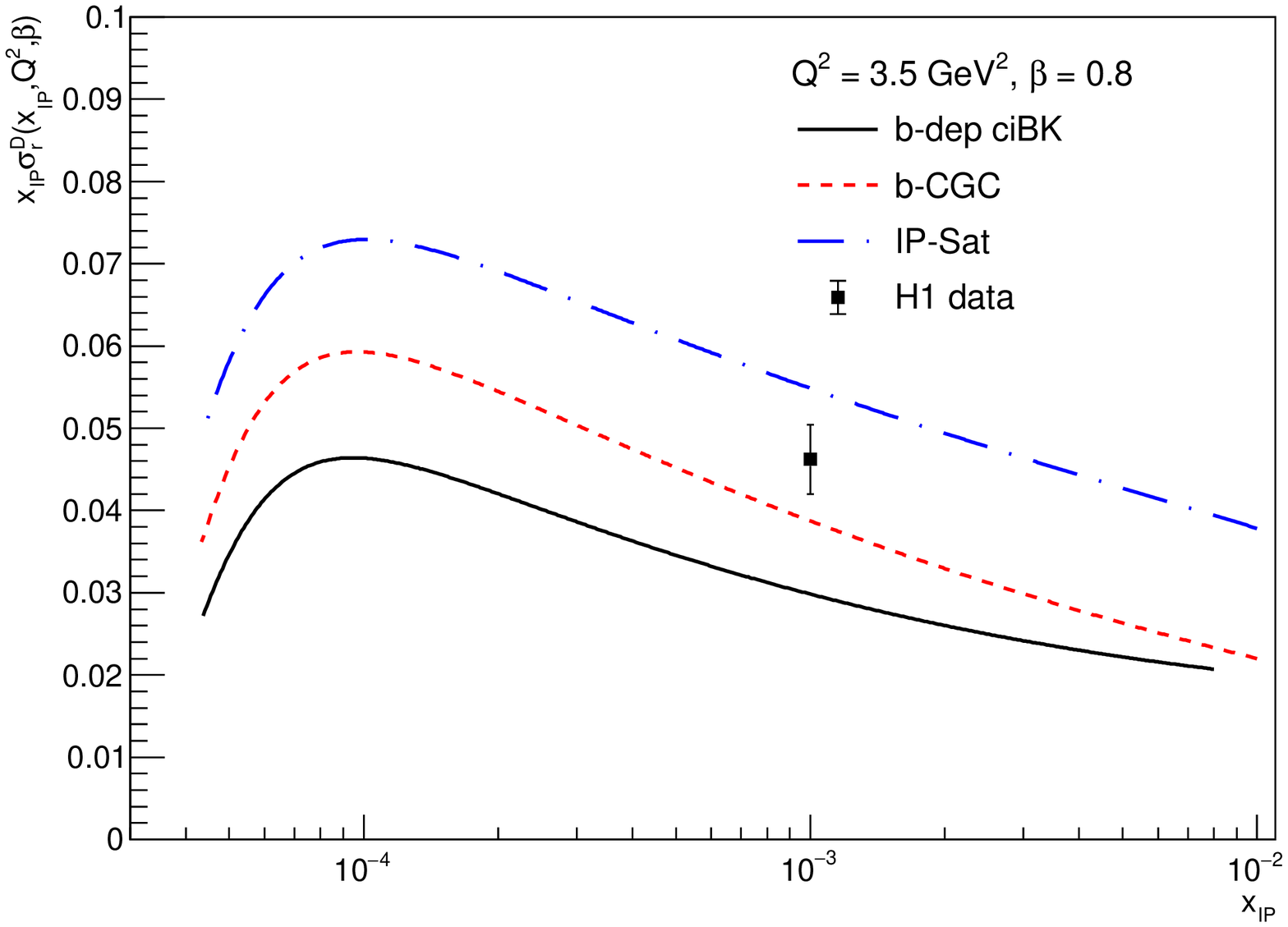}} \\ 
 {\includegraphics[width=0.33\textwidth]{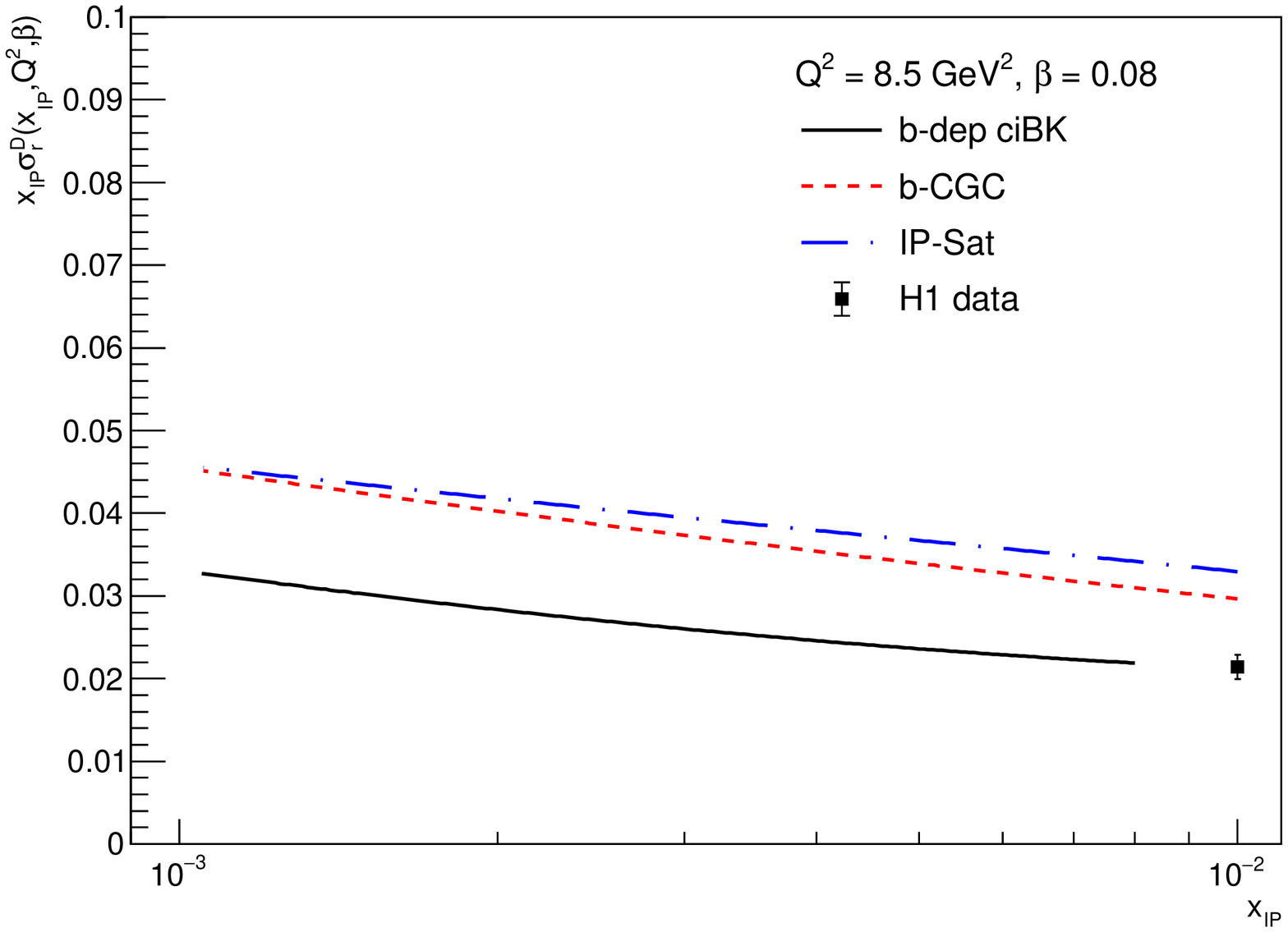}} & 
{\includegraphics[width=0.33\textwidth]{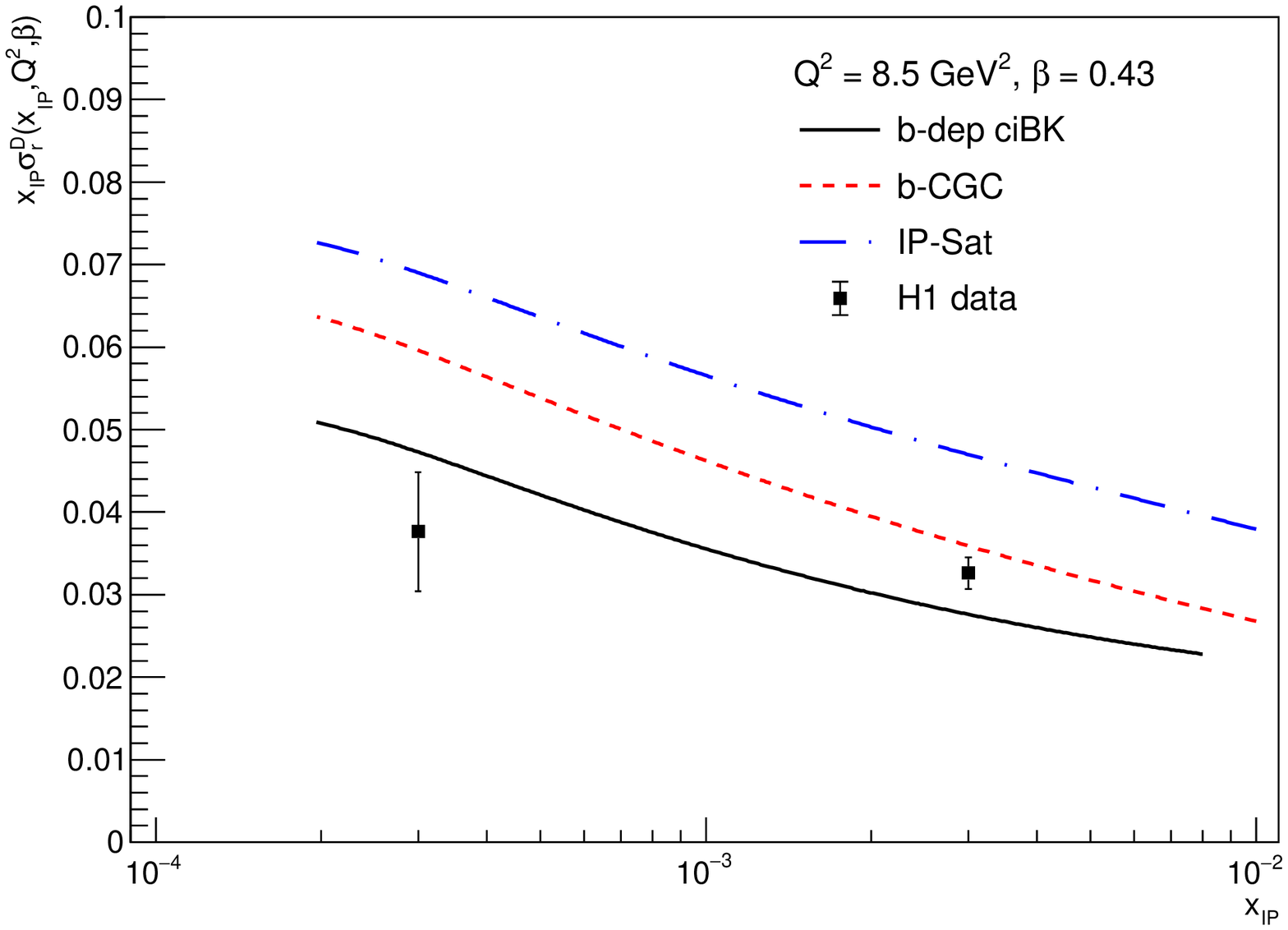}} &
{\includegraphics[width=0.33\textwidth]{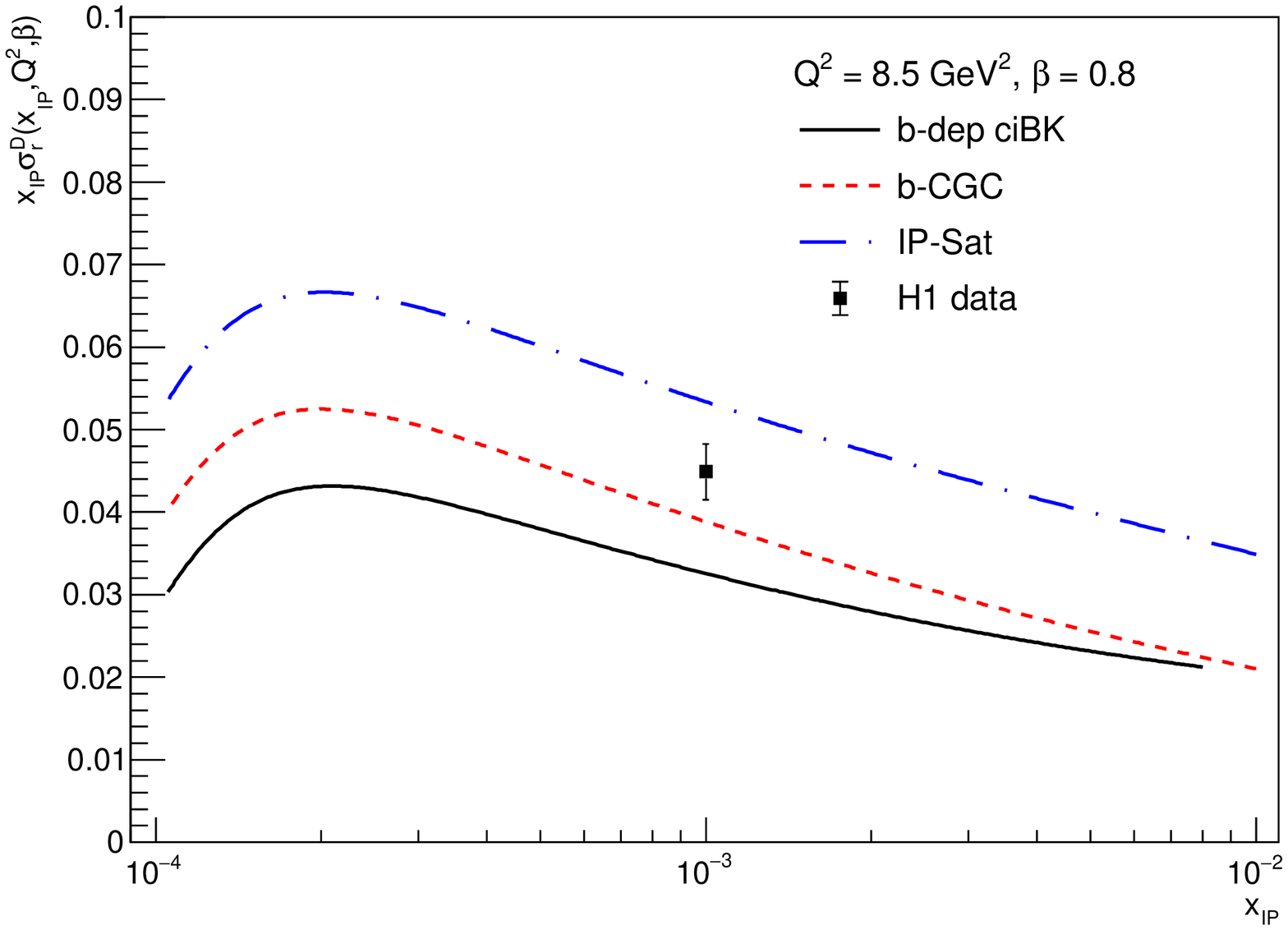}}\\
\end{tabular}                                                                                                                       
\caption{Predictions for the $x_{\pom}$-dependence of  the reduced diffractive cross section $\sigma^{D(3)}_{r}(x_{\pom},\beta,Q^{2})$ at different combinations of the photon virtuality and $\beta$, considering the b-CGC, IP-Sat, and b-dep ciBK models for the dipole-proton scattering amplitude. Data from H1 are presented for comparison \cite{Aaron:2012ad}.}
\label{fig:f2d3_data}
\end{figure}

In Fig.~\ref{fig:f2d3_data}, we present the predictions for the $x_{\pom}$-dependence of the reduced diffractive cross section $\sigma^{D(3)}_{r}(x_{\pom},\beta,Q^{2})$ at different combinations of the photon virtuality and  values of $\beta$, considering the b-CGC, IP-Sat, and b-dep ciBK models for the dipole-proton scattering amplitude. The data from H1 are presented for comparison \cite{Aaron:2012ad}. The $x_{\pom}$-dependencies of the predictions are similar. However, one has that the normalization of the b-dep ciBK prediction is smaller than the b-CGC and IP-Sat one, in agreement with the results derived for the diffractive structure function. We observe that the data for medium and small $\beta$ is better described by the b-dep ciBK model, but this model underestimates the data at large $\beta$ values. The opposite situation occurs for the other two models. Unfortunately, due to the scarcity of data, a more definitive conclusion is not possible. However, our results indicate that a future analysis of this observable, to be performed in the EIC and LHeC, will be very useful to constrain the underlying assumptions about the QCD dynamics.


As discussed in previous sections, the impact of  nonlinear effects is expected to be enhanced in $eA$ collisions and, as demonstrated in Fig.~\ref{fig:ratio}, the contribution from diffractive events is larger in the nuclear case. Such results motivate the study of the diffractive structure functions $F_2^{D (3)}$ and the reduced diffractive cross sections $\sigma^{D(3)}_{r}(x_{\pom},\beta,Q^{2})$ for different nuclei. In our analysis, we estimate the behavior of the ratio between these quantities for nuclei and the associated predictions for $ep$ collisions re-scaled by the atomic number $A$. If the nuclear effects are negligible, such ratios are equal to unity. Therefore, the study of these ratios for different nuclei allows us to investigate the impact of nonlinear corrections in the kinematic range that will be probed in the future electron-ion colliders. 

\begin{figure}[t]
\begin{tabular}{ccc}
{\includegraphics[width=0.33\textwidth]{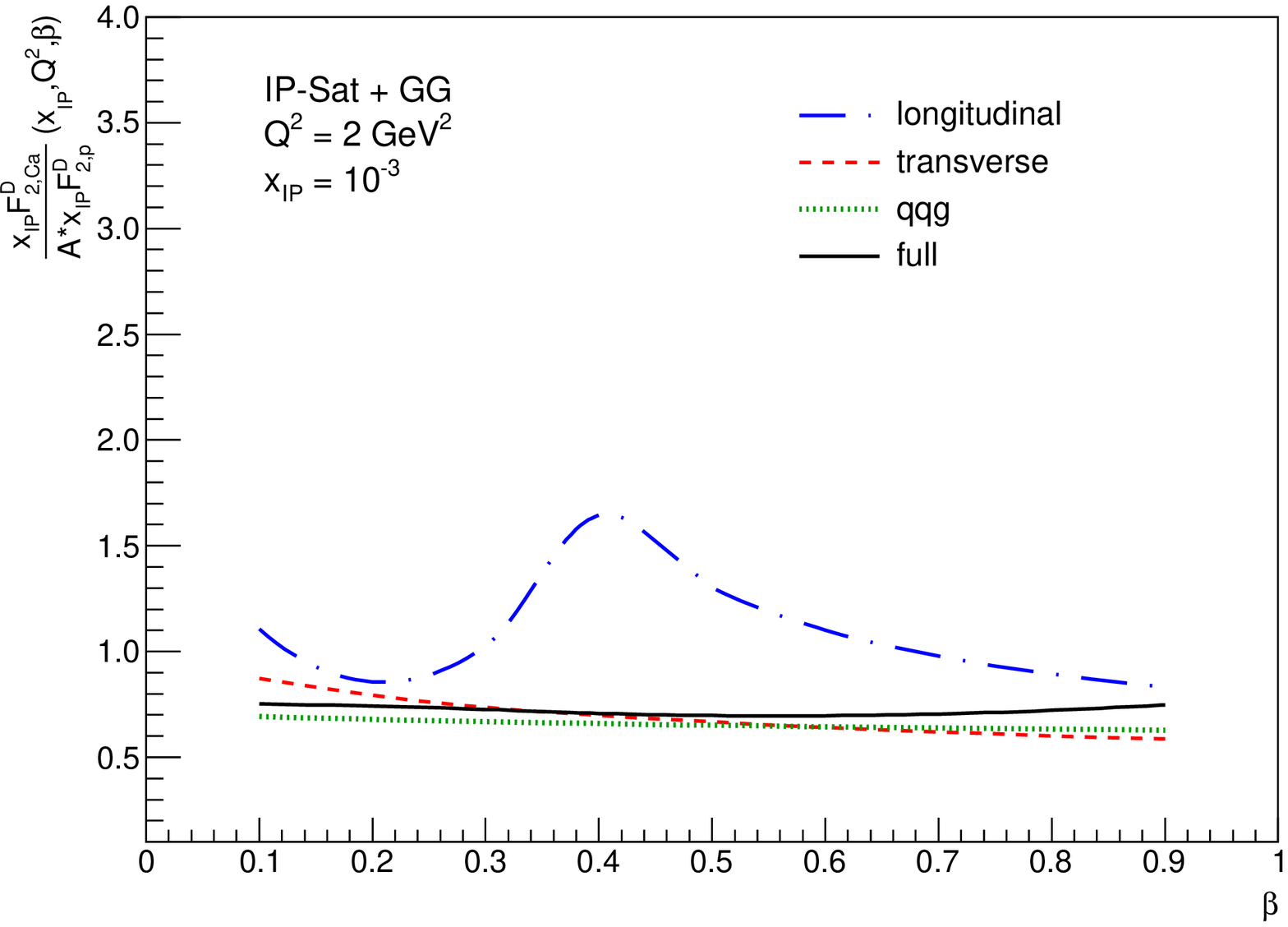}} &
 {\includegraphics[width=0.33\textwidth]{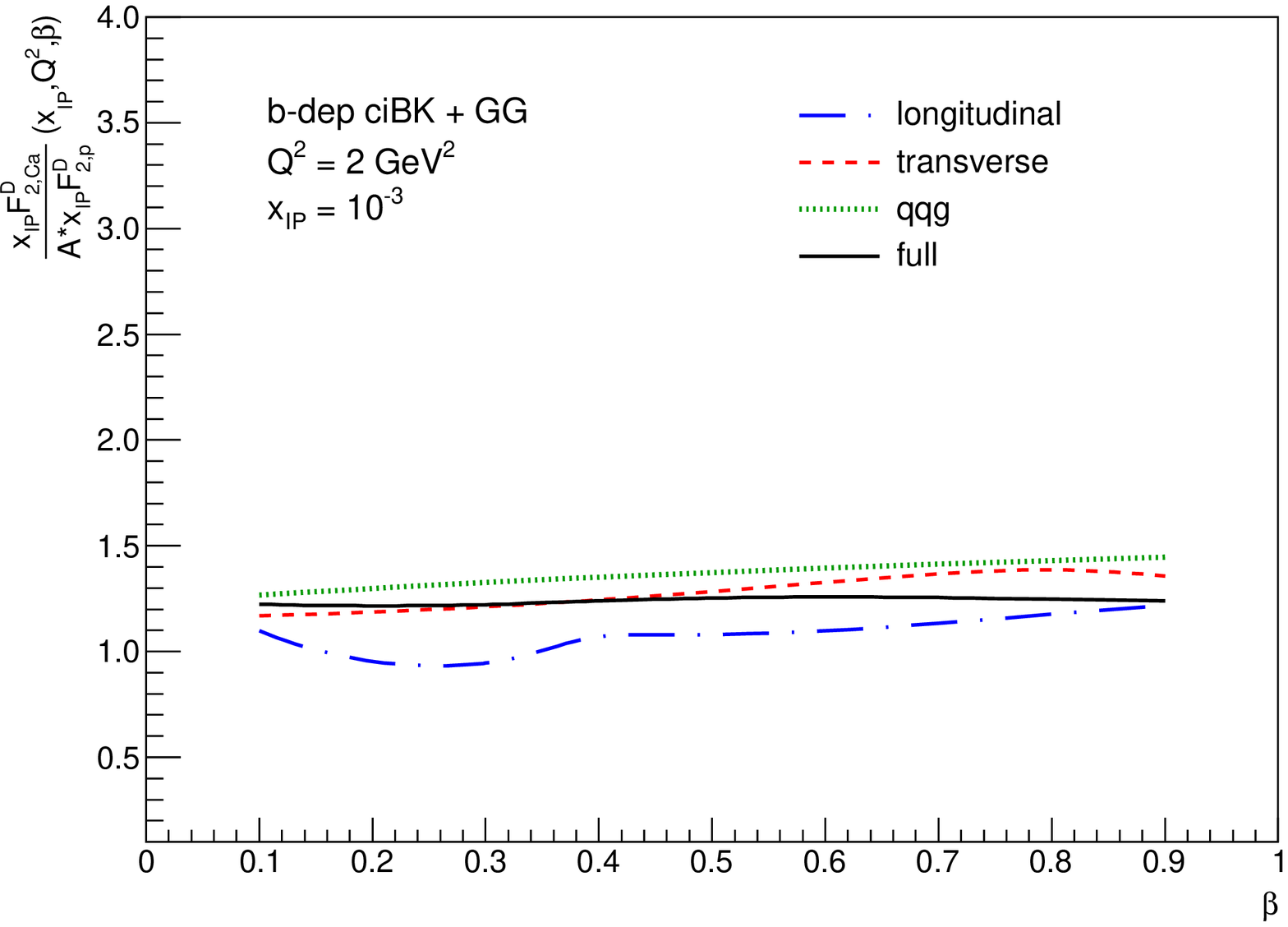}} & 
{\includegraphics[width=0.33\textwidth]{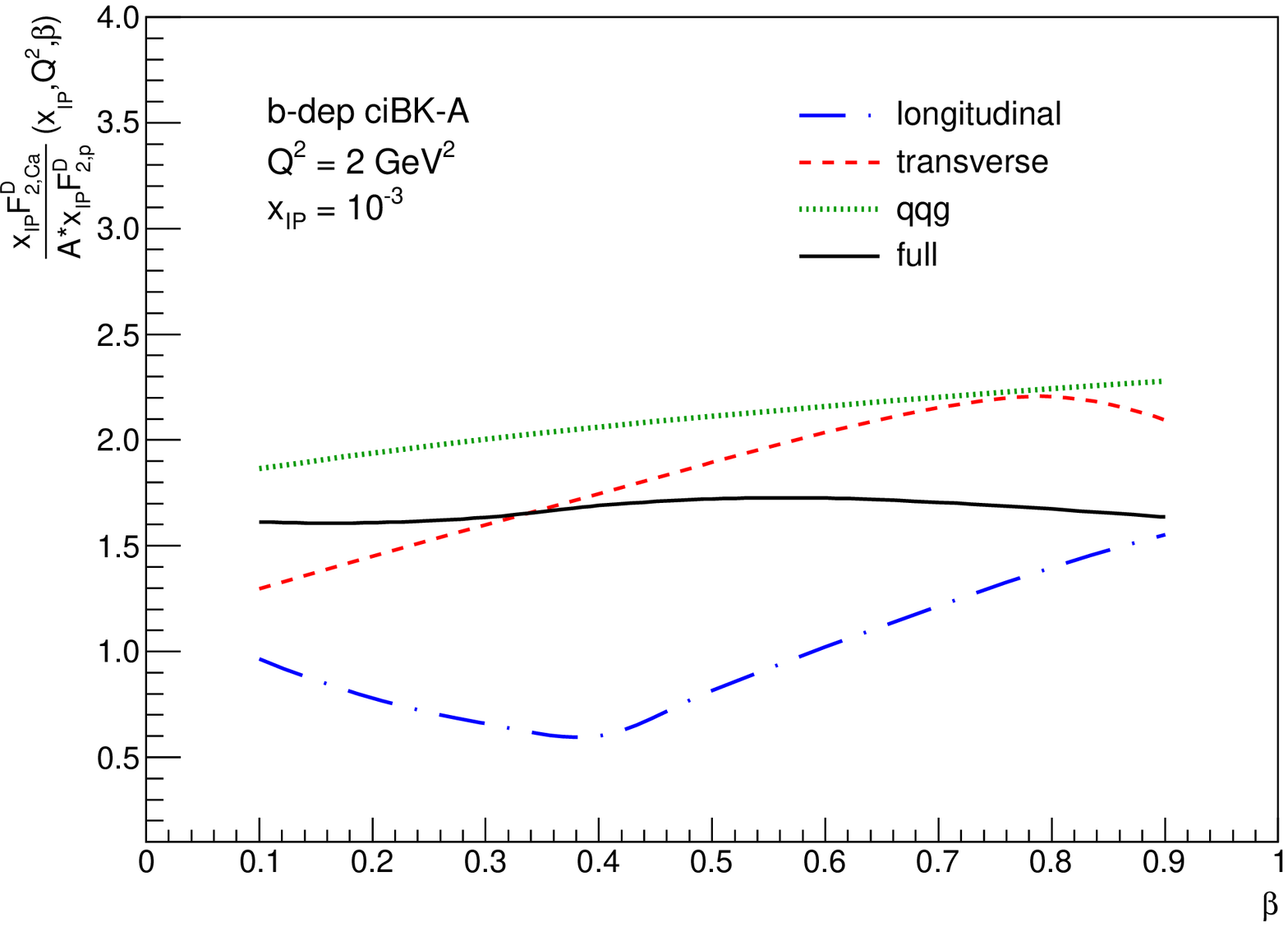}} \\
{\includegraphics[width=0.33\textwidth]{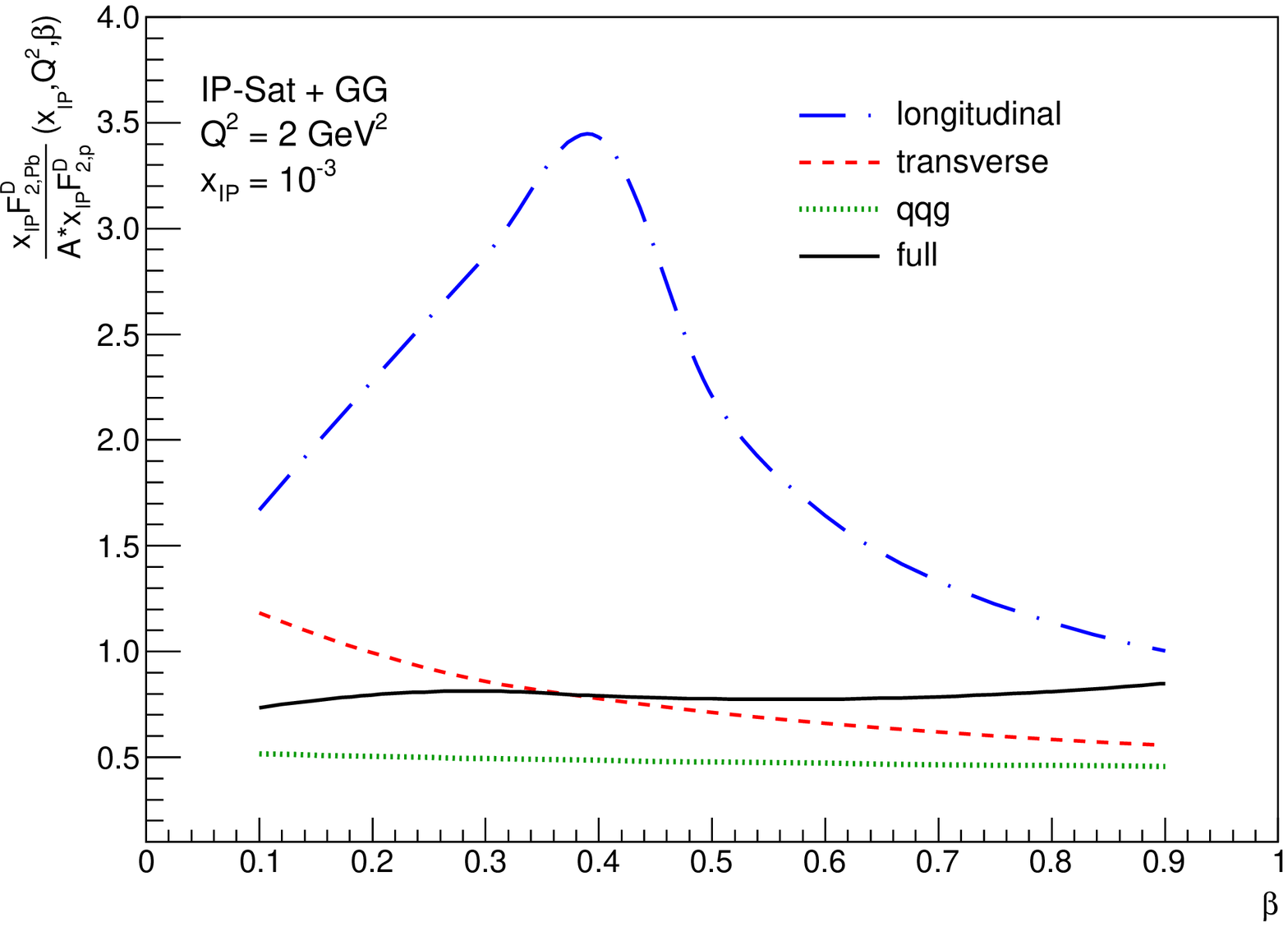}} & {\includegraphics[width=0.33\textwidth]{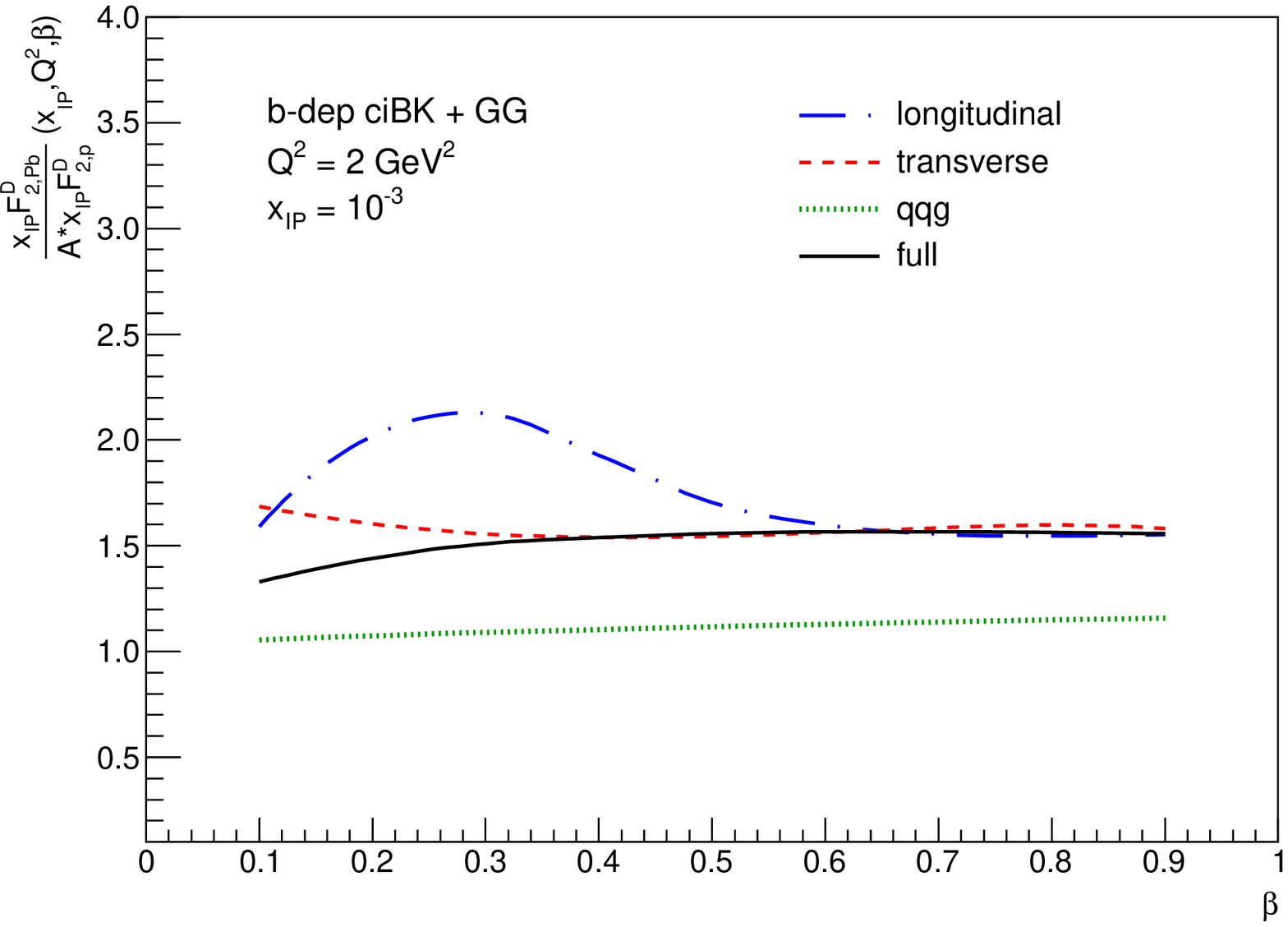}} & 
{\includegraphics[width=0.33\textwidth]{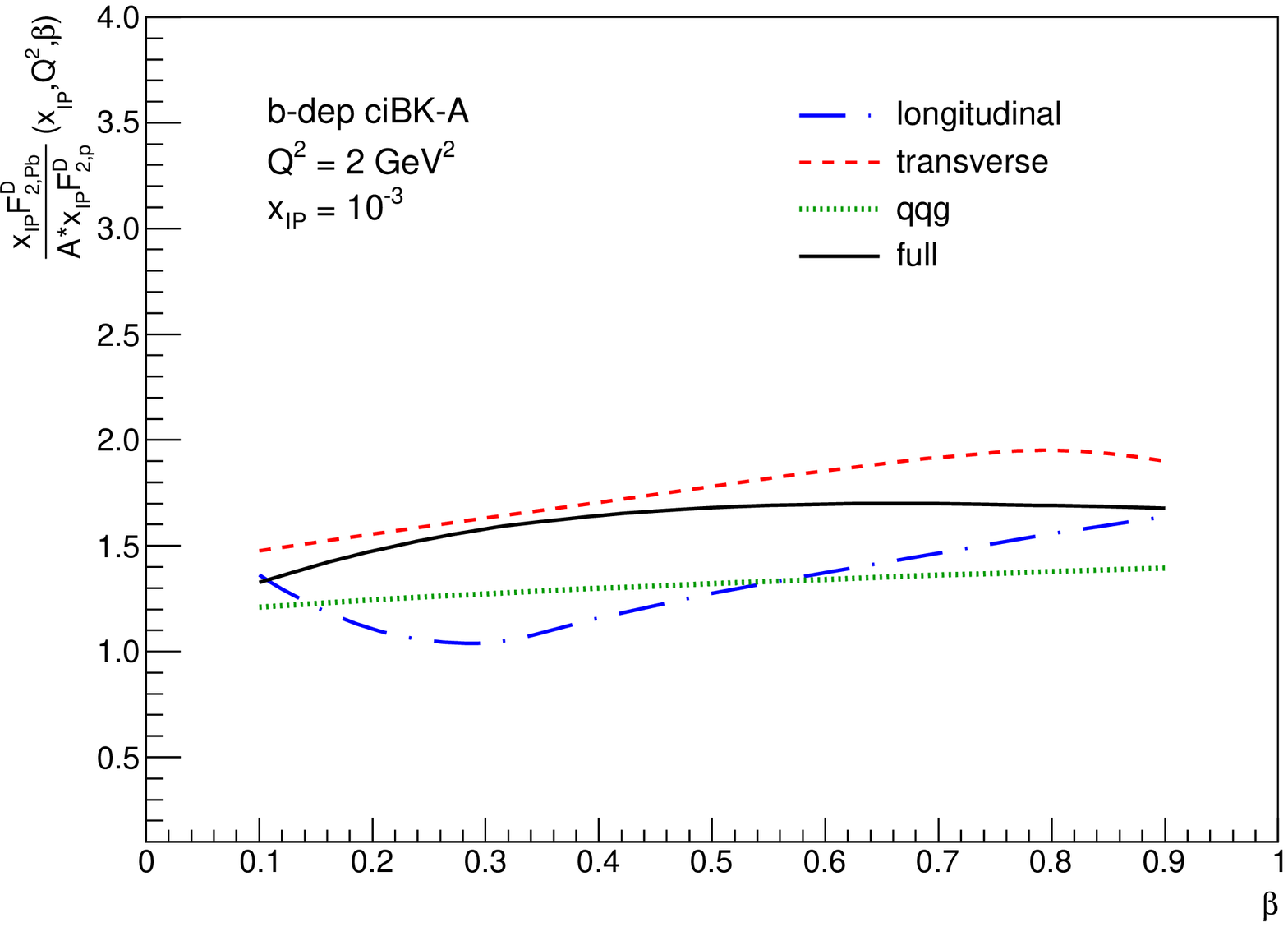}} \\
\end{tabular}                                                                                                                       
\caption{Predictions for the $\beta$-dependence of the ratio $F_2^{D (3),A}/AF_2^{D (3),p}$ for $A = 40$ (upper panels) and $A = 208$ (lower panels) considering the IP-Sat + GG (left panels), b-dep ciBK + GG (middle panels) and b-dep ciBK-A (right panels) models for the dipole-nucleus scattering amplitude. Predictions for the distinct components are presented separately.}
\label{fig:f2dnuc_beta}
\end{figure}

Initially, in Fig.~\ref{fig:f2dnuc_beta} we present our predictions for the $\beta$-dependence of the ratio $F_2^{D (3),A}/AF_2^{D (3),p}$ for $A = 40$ (upper panels) and $A = 208$ (lower panels), considering the IP-Sat + GG (left panels), b-dep ciBK + GG (middle panels) and b-dep ciBK-A (right panels) models for the dipole-nucleus scattering amplitude. We set $Q^2 = 2$ GeV$^2$ and $x_{\pom} = 10^{-3}$, these kinematic values can be potentially studied in the future $eA$ colliders. Moreover, we present the predictions for the different components that contribute to the diffractive structure function. Our results for the IP-Sat + GG model agree with those presented in Ref.~\cite{Kowalski_prc} using the same approach. However, we show that the predictions are strongly dependent on the modeling of the dipole-target scattering amplitude. In particular, b-dep ciBK models predict a nuclear enhancement of all components that contribute to the diffractive structure function, which is mainly associated with the smaller normalization of the diffractive structure function of the proton in comparison to the IP-Sat model. We have that the b-dep ciBK-A predictions for the sum of the components  (denoted full in the figures) are similar for the two nuclei, which is associated to the strong suppression of the $q\bar{q}g$ component with increasing atomic number $A$. Such suppression is also predicted in Ref.~\cite{Kowalski_prc}.

\begin{figure}[t]
\begin{tabular}{ccc}
 {\includegraphics[width=0.33\textwidth]{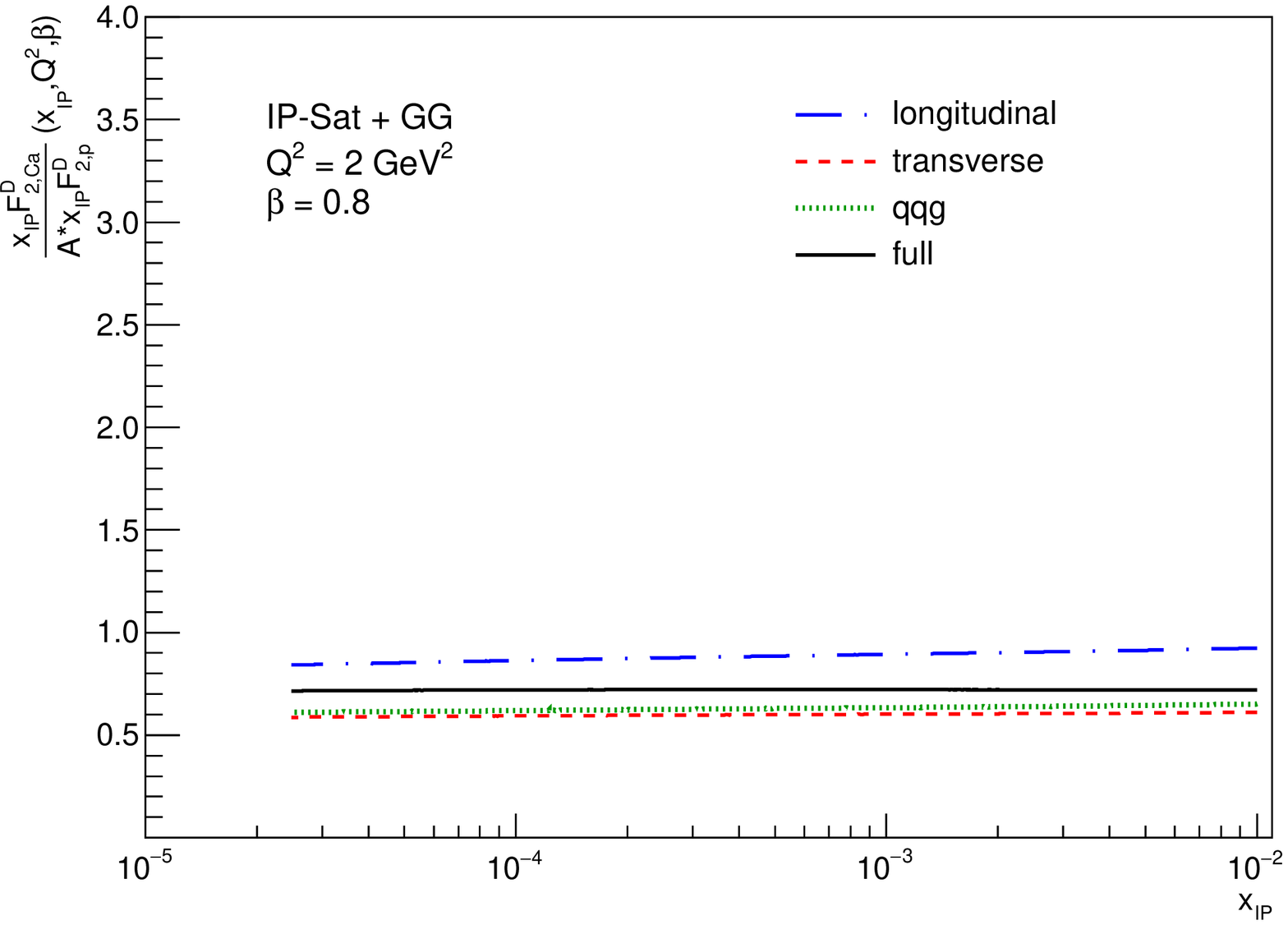}} &  {\includegraphics[width=0.33\textwidth]{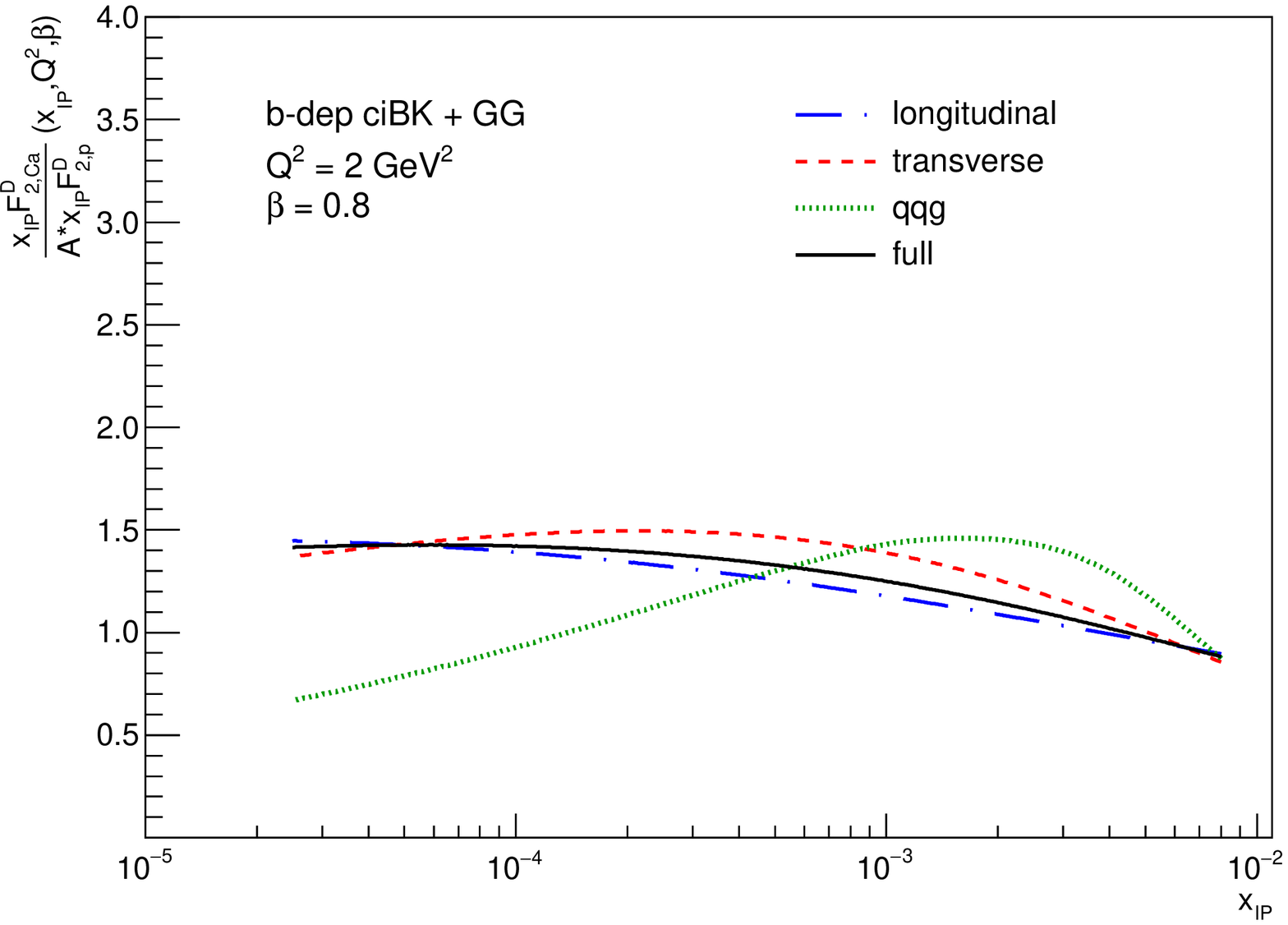}} & 
{\includegraphics[width=0.33\textwidth]{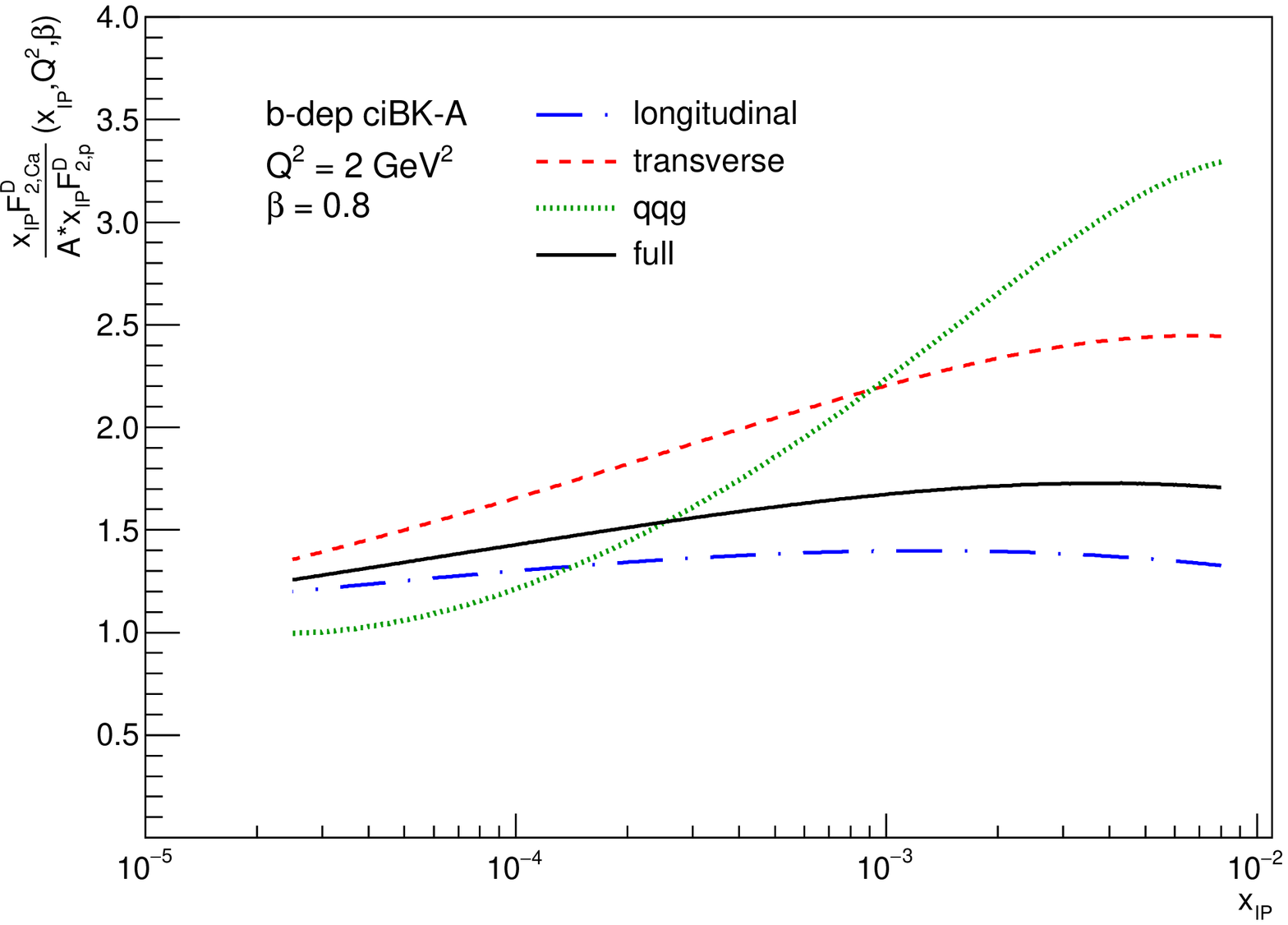}} \\
 {\includegraphics[width=0.33\textwidth]{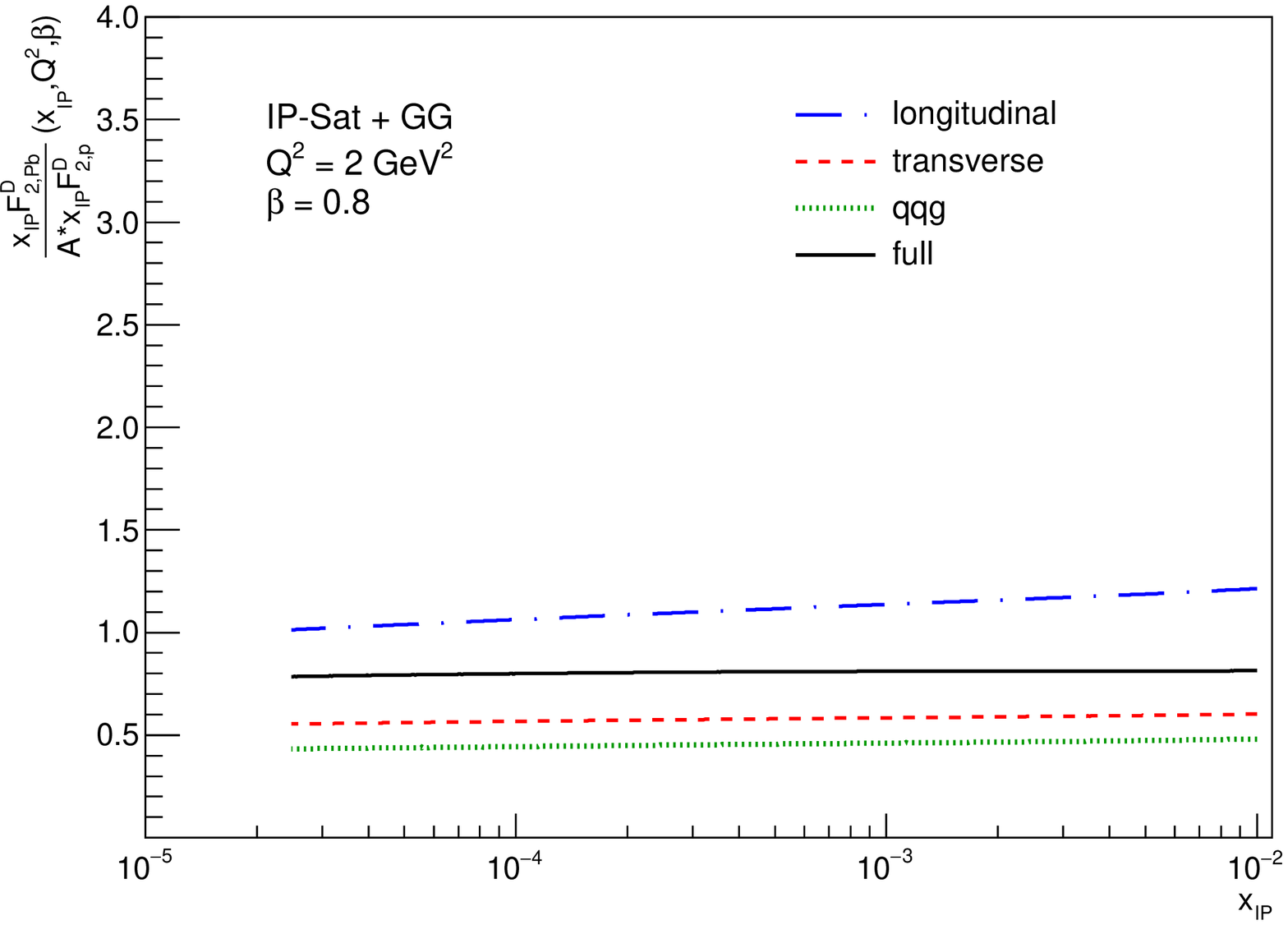}} &  {\includegraphics[width=0.33\textwidth]{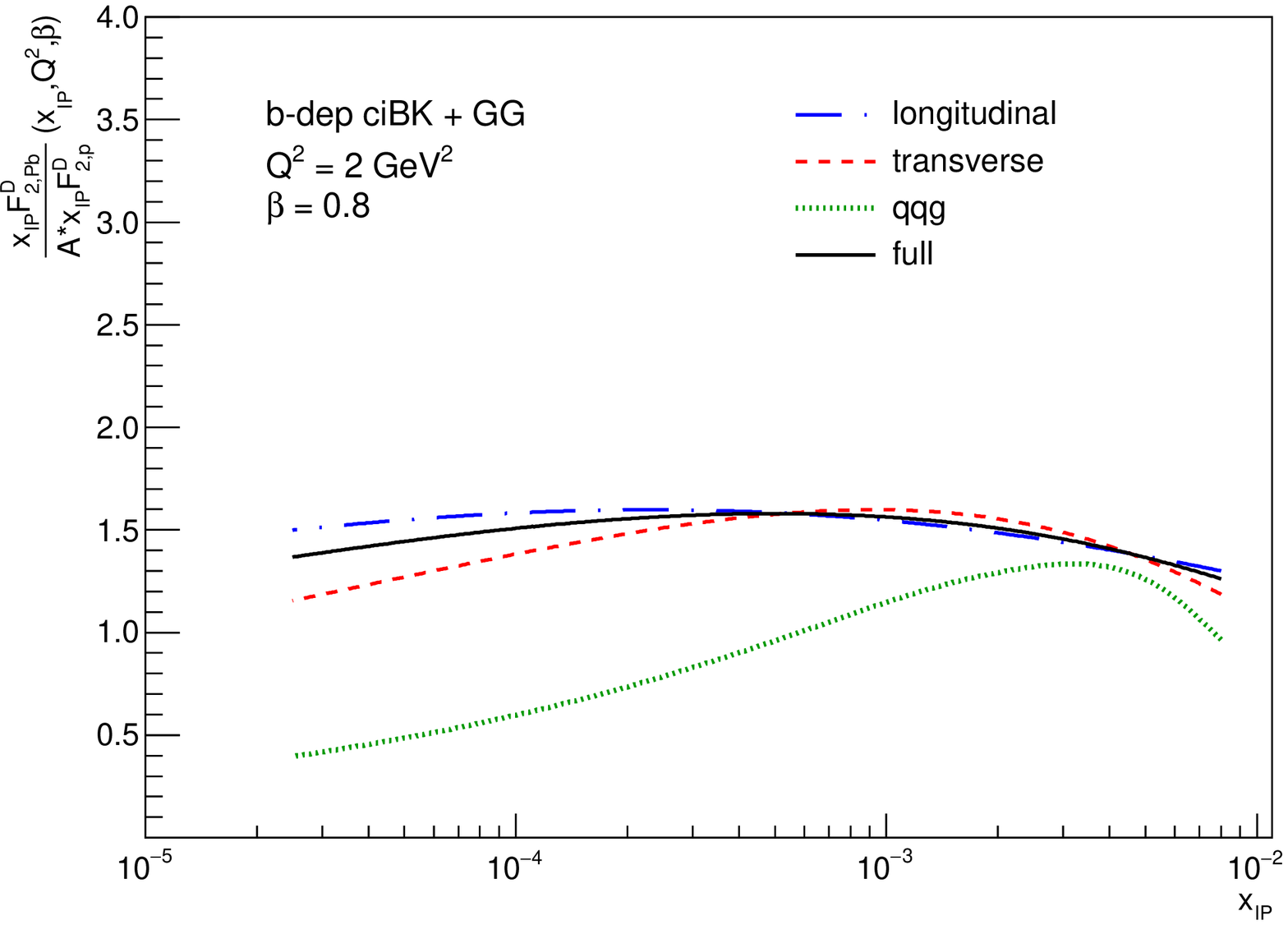}} & 
{\includegraphics[width=0.33\textwidth]{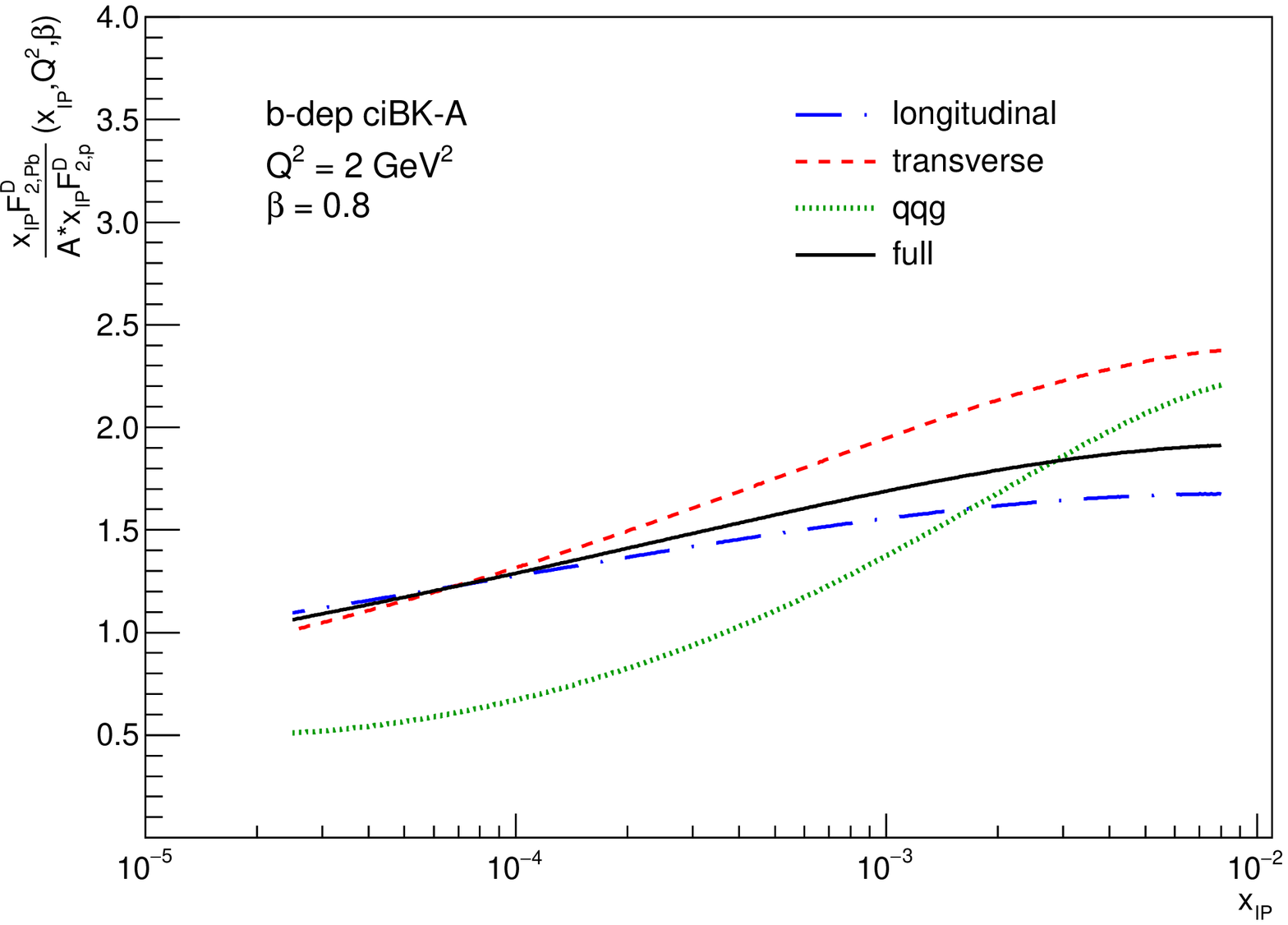}} \\
\end{tabular}                                                                                                                       
\caption{Predictions for the $x_{\pom}$-dependence of the ratio $F_2^{D (3),A}/AF_2^{D (3),p}$ for $A = 40$ (upper panels) and $A = 208$ (lower panels) considering the IP-Sat + GG (left panels), b-dep ciBK + GG (middle panels) and b-dep ciBK-A (right panels) models for the dipole-nucleus scattering amplitude. Predictions for the distinct components are presented separately.}
\label{fig:f2dnuc_xpom}
\end{figure}

\begin{figure}[t]
\begin{tabular}{cc}
 {\includegraphics[width=0.5\textwidth]{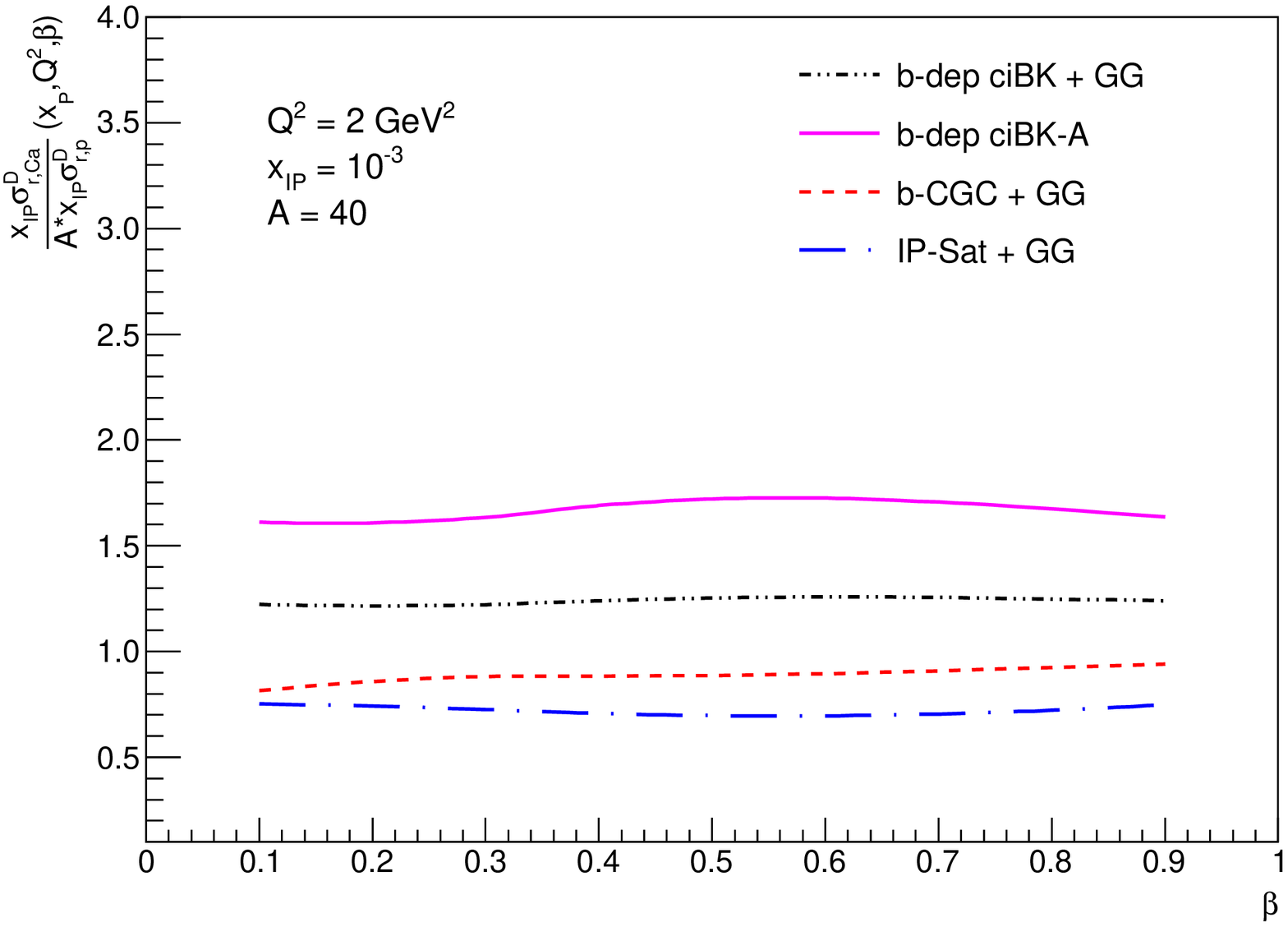}} & 
{\includegraphics[width=0.5\textwidth]{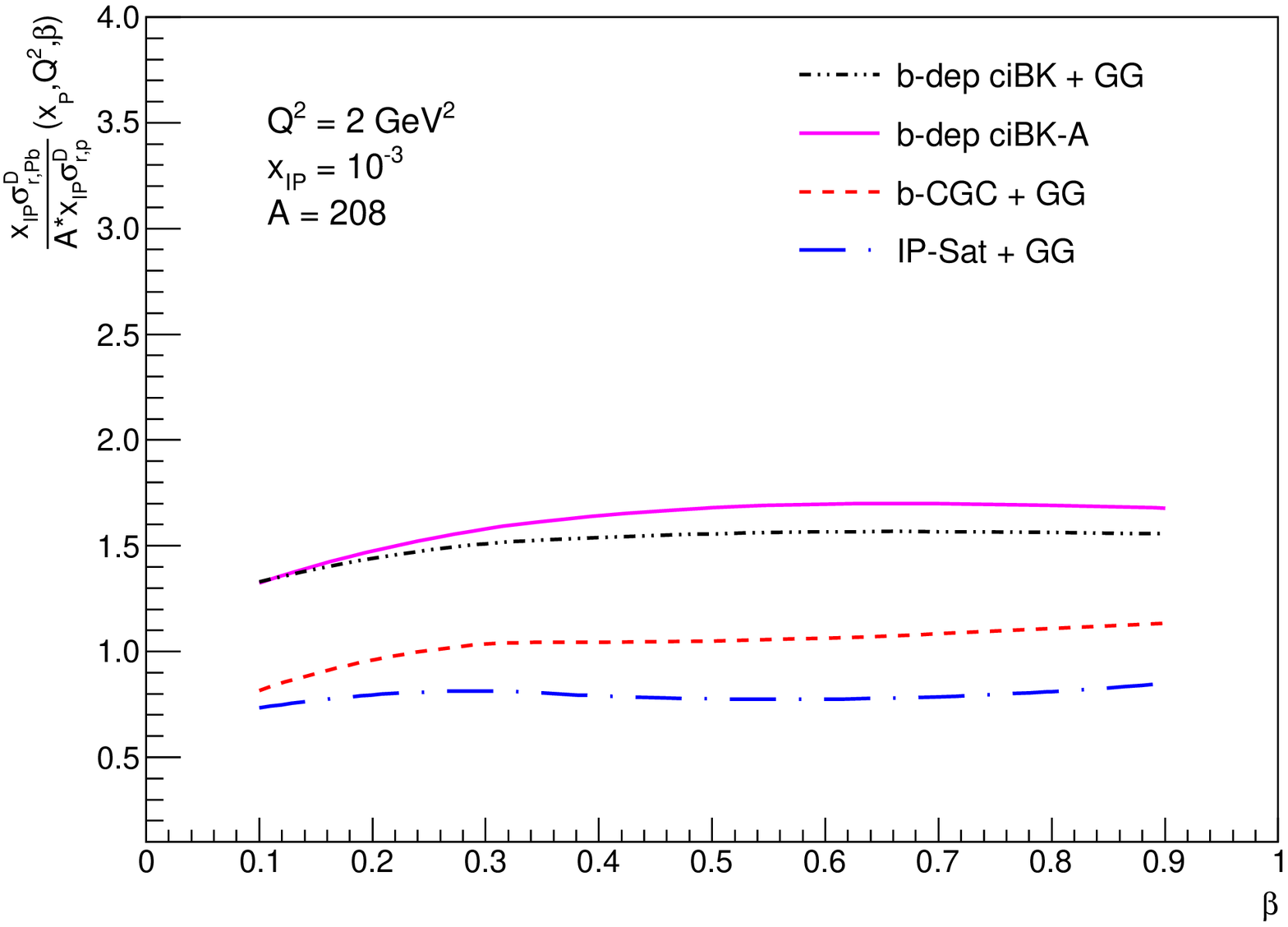}} \\
 {\includegraphics[width=0.5\textwidth]{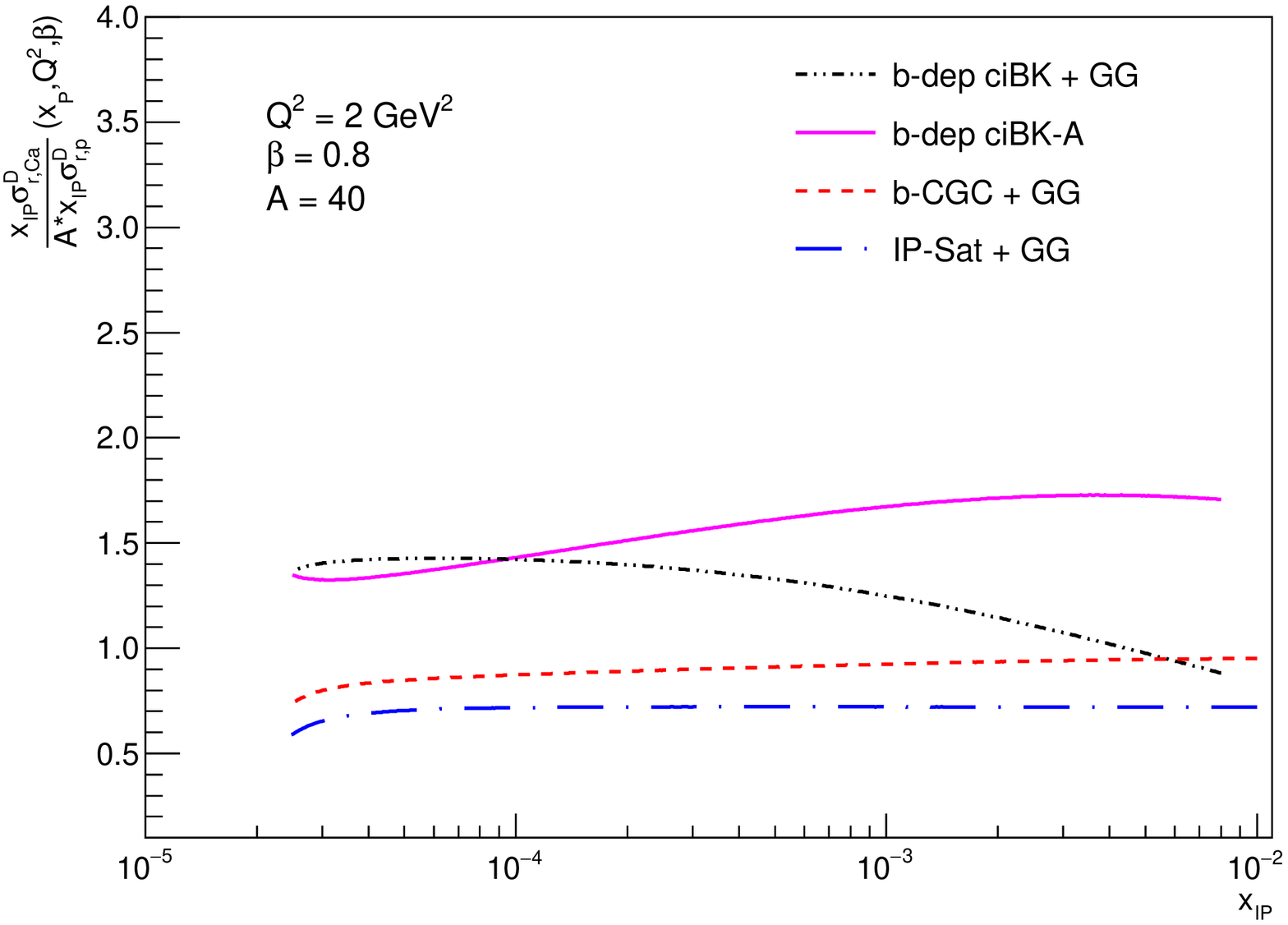}} & 
{\includegraphics[width=0.5\textwidth]{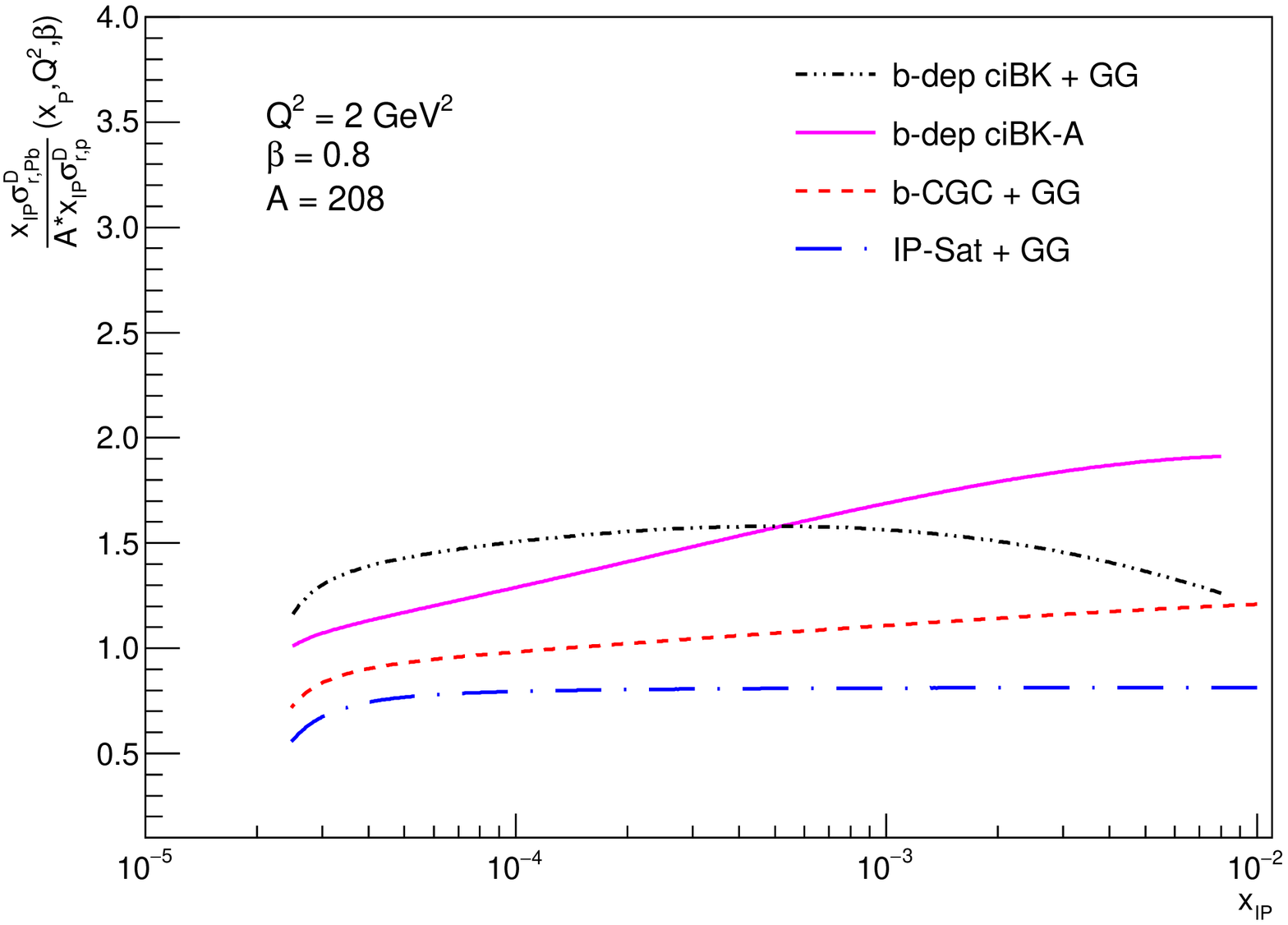}} \\
\end{tabular}                                                                                                                       
\caption{Predictions for the $\beta$ (upper panels) and $x_{\pom}$ (lower panels) dependencies of the ratio $\sigma^{D(3),A}_{r}(x_{\pom},\beta,Q^{2})/ A \cdot \sigma^{D(3),p}_{r}(x_{\pom},\beta,Q^{2})$ for $A = 40$ (left panels) and $A = 208$ (right panels) considering the different models for the dipole-nucleus scattering amplitude.}
\label{fig:Sigred}
\end{figure}

In Fig.~\ref{fig:f2dnuc_xpom}, we present our predictions for the $x_{\pom}$-dependence of the ratio $F_2^{D (3),A}/AF_2^{D (3),p}$ for $A = 40$ (upper panels) and $A = 208$ (lower panels) assuming  $Q^2 = 2$ GeV$^2$ and $\beta = 0.8$. For this value of $\beta$, the diffractive structure is mostly dominated by the longitudinal component, with the associated ratio being slightly larger for a heavier nuclei, in agreement with the results obtained in Ref.~\cite{Kowalski_prc}. However, our results demonstrate that the $x_{\pom}$-dependence of the ratio is dependent on the model used to describe the dipole-nucleus interaction. Finally, in Fig.~\ref{fig:Sigred}, we present our predictions for the $\beta$ (upper panels) and $x_{\pom}$ (lower panels) dependencies of the ratio $\sigma^{D(3),A}_{r}(x_{\pom},\beta,Q^{2})/ A \cdot \sigma^{D(3),p}_{r}(x_{\pom},\beta,Q^{2})$ for $A = 40$ (left panels) and $A = 208$ (right panels), considering the different models for the dipole-nucleus scattering amplitude. Our results indicate that a future experimental analysis of this observable for two distinct nuclei will be very useful, since the normalization and dependencies on $\beta$ and $x_{\pom}$ are strongly dependent on the description of the QCD dynamics. In particular, one can observe that the predictions derived using the impact-parameter dependent BK equation are larger than those obtained using the phenomenological models.

\section{Summary}
\label{sec:conc}

Future electron-ion collisions will allow us to study the high-gluon density regime of QCD, where the contribution of nonlinear (saturation) effects are expected to determine the behavior of the inclusive, diffractive and exclusive observables. In this paper, we have investigated the impact of these effects on diffractive observables. In particular, we have presented, for the first time, the predictions for the diffractive structure functions and reduced diffractive cross sections derived using the solution of the impact-parameter dependent Balitsky-Kovchegov equation for the dipole-target scattering amplitude. We have presented the predictions for $ep$ and $eA$ collisions, considering the kinematic ranges that will be probed by the EIC, LHeC and FCC-$eh$. It has been demonstrated that the contribution of the diffractive events increases with the atomic number, being of the order of 20\% for $ePb$ collisions, with this prediction being almost independent on the modeling of the dipole-nucleus interaction. Our results for $ep$ collisions indicate that the BK equation satisfactorily describes the data and implies a smaller normalization for the reduced cross section in comparison to the phenomenological models based on the CGC physics. For $eA$ collisions, we have shown that a future experimental analysis of the diffractive observables will be useful to improve our understanding of QCD dynamics at high parton densities.

\begin{acknowledgments}
VPG would like to thank the members of the Czech Technical University in Prague for the warm hospitality during the beginning of this project. VPG was partially financed by the Brazilian funding agencies CNPq, FAPERGS and INCT-FNA (processes number 464898/2014-5). This work has been partially supported by the grant LTC17038 of the INTER-EXCELLENCE program at the Ministry of Education, Youth and Sports of the Czech Republic, by the grant 18-07846Y of the Czech Science Foundation (GACR) and by the Centre of Advanced Applied Sciences with the number:CZ.02.1.01/0.0/0.0/16-019/0000778. The Centre of Advanced Applied Sciences is co-financed by the European Union.
\end{acknowledgments}

\hspace{1.0cm}

\end{document}